\documentclass[twocolumn,showpacs,preprintnumbers,amsmath,amssymb]{revtex4}
\usepackage{graphicx}
\usepackage{dcolumn}
\usepackage{bm}

\begin{document}
\title{Soliton  fractional charge  of   disordered      graphene  nanoribbon}
\author{Y. H. Jeong$^{1}$, S.-R. Eric Yang$^{1}$\footnote{Corresponding author, eyang812@gmail.com}, M.-C. Cha$^{2}$}
\affiliation{ $^{1}$Department of Physics, Korea  University, Seoul, Korea\\ 
$^{2}$Department of Photonics and Nanoelectronics, Hanyang  University, Ansan, Korea\\}

\begin{abstract}
We investigate  the  properties   of the gap-edge states of half-filled interacting disordered   zigzag graphene nanoribbons, and find that the midgap states can display  a quantized fractional charge of  $1/2$. These gap-edge states  can be represented by  topological kinks with  their  site probability distribution divided between the left and right zigzag edges with different chiralities.  In addition, there are numerous spin-split gap-edge  states, similar to those in a Mott-Anderson insulator.
\end{abstract}

\maketitle

\section{Introduction}

First principles and Hubbard model calculations\cite{Waka1, Son}, along with the investigation of the topological Zak phase\cite{Zak, Jeong1}, have shown that spin-up integer charges are localized on one zigzag edge of a graphene nanoribbon (GNR), while  spin-down charges are localized on the opposite zigzag edge. The repulsive interaction between electrons is crucial for this effect, as it reduces the double occupancy of the spin-up and -down electrons.  Although numerous one-dimensional insulators, such as  polyacetylene, spin chains, and Kondo insulators,  have   fractional charges on their boundaries\cite{Kit,Jack1,Sol,Ng,Piers}, no fractional charge has been found on the well-separated zigzag   edges of an undoped GNR; instead, only integer charges have been identified\cite{Waka1,Son, Jeong1, Yang}.    
Antiferromagnetic coupling between the well-separated zigzag edges may be responsible for suppression of the fractional edge charge formation.

However, GNRs can support solitonic states, which are usually associated with a fractional charge\cite{Jack1,Sol}.  One of the key features of a soliton  is that, for each of  its spin states, half its spectral weight originates from the conduction band and the other half from the valence band.  Application of tensile strain locally in  GNRs may create a  soliton, with its site pseudospin\cite{pseudo}  connecting  two zigzag edges of opposite chirality\cite{Jeong0}.
In a doped zigzag graphene nanoribbon (ZGNR), the system prefers energetically to create magnetic domains and accommodate the extra electrons in the interface solitons\cite{Luis}; such a domain wall is analogous to that of polyacetylene connecting two different dimerized phases\cite{Sol}. A fractional charge is well-defined when its spatial overlap with other fractional charges is small. However, measurement of isolated and well-separated fractional charges in a system to which a tensile strain is applied or numerous electrons are added is not straightforward.

In graphene nano systems, gap states are localized and reside on the zigzag edges. They play a central role in determining the edge and magnetic properties of  graphene nano systems\cite{Ross}  (in the following, we call these states  ``gap-edge states").
An impurity can have a significant effect on  the probability density redistribution of an edge state over   the left and right zigzag edges of a ZGNR\cite{Park}.

In this  work, we investigate the influence  of disorder on the gap-edge states  in the presence of antiferromagnetic coupling.
In particular, we investigate whether well-separated  fractional  zigzag edge charges may appear in the presence of disorder  in  ZGNRs.  
To investigate  the formation of  a fractional edge charge of $1/2$, it is necessary to include 
both electron-electron interactions and disorder in a self-consistent manner (the elementary charge is set to $e=1$). Here, we perform such a calculation for an undoped and disordered ZGNR, where
the number of gap states increases with increasing  zigzag edge length.   We find that disorder weakens the antiferromagnetic coupling between the zigzag edges and, hence, stabilizes a solitonic state with fractional charges.  Among numerous gap states,  only midgap states    (energy $E \approx 0)$  contribute to the fractional  charge of $1/2$  in the weak disorder  regime, where the disorder strength  is smaller than the on-site repulsion.       The  formation of a well-defined  fractional charge  originates from  the left and right edge states coupled by a short-ranged disorder potential, forming bonding or antibonding states.  
We find that the charge fractionalization of a midgap state is more robust for short-ranged disorder potentials than for  long-ranged potentials.  The site pseudospin  of the midgap states exhibits a topological kink-like property.  It connects  the left and right zigzag edges with different directions of  magnetization,  analogous to a domain wall soliton in polyacetylene that connects two dimerized phases.  Such a   soliton charge  is topologically protected  against weak disorder.  In addition, we find that several edge states become spin-split and singly occupied, as in a Mott-Anderson insulator\cite{Dob}.

This paper is organized as follows. In Sec.~2, we describe various types of zigzag edge (gap) states, which we classify as  Types  I, II, and III.
In Sec.~3, we describe our model and  the mean field Hamiltonian. The fractional charge is defined in Sec.~4 using the eigenstates of the Hamiltonian.  In Sec.~5,  the site probability  distributions of the Type-I, -II, and -III gap states and their localization properties  are computed.
The non-integer charge distributions of the gap states are investigated as  functions of  their energy in Sec.~6.  In Sec.~7, we investigate the relationship between variance of  the quantized value of the fractional charge and the impurity range.  These results are used to show the quantization of the charge fractionalization of the midgap states.    A summary and discussion are given in Sec.~8.

\section{Types of zigzag edge state }

Here, we describe basic and well-known properties of zigzag edge states and define notation, which will help us better explain the results reported in the sections below.    

A rectangular GNR  has two zigzag edges and two armchair edges. When the length of the zigzag edges is longer than that of the armchair edges, a ZGNR is realized; in the inverse case, an armchair graphene nanoribbon is realized.   A periodic ZGNR with a bandstructure has only two zigzag edges and no armchair edges. Furthermore, electron-electron interactions generate an excitation gap\cite{Yang} and induce antiferromagnetism between the opposite edges of a ZGNR\cite{Waka1}.

\begin {figure}[!hbpt]
\begin{center}
\includegraphics[width=0.3\textwidth]{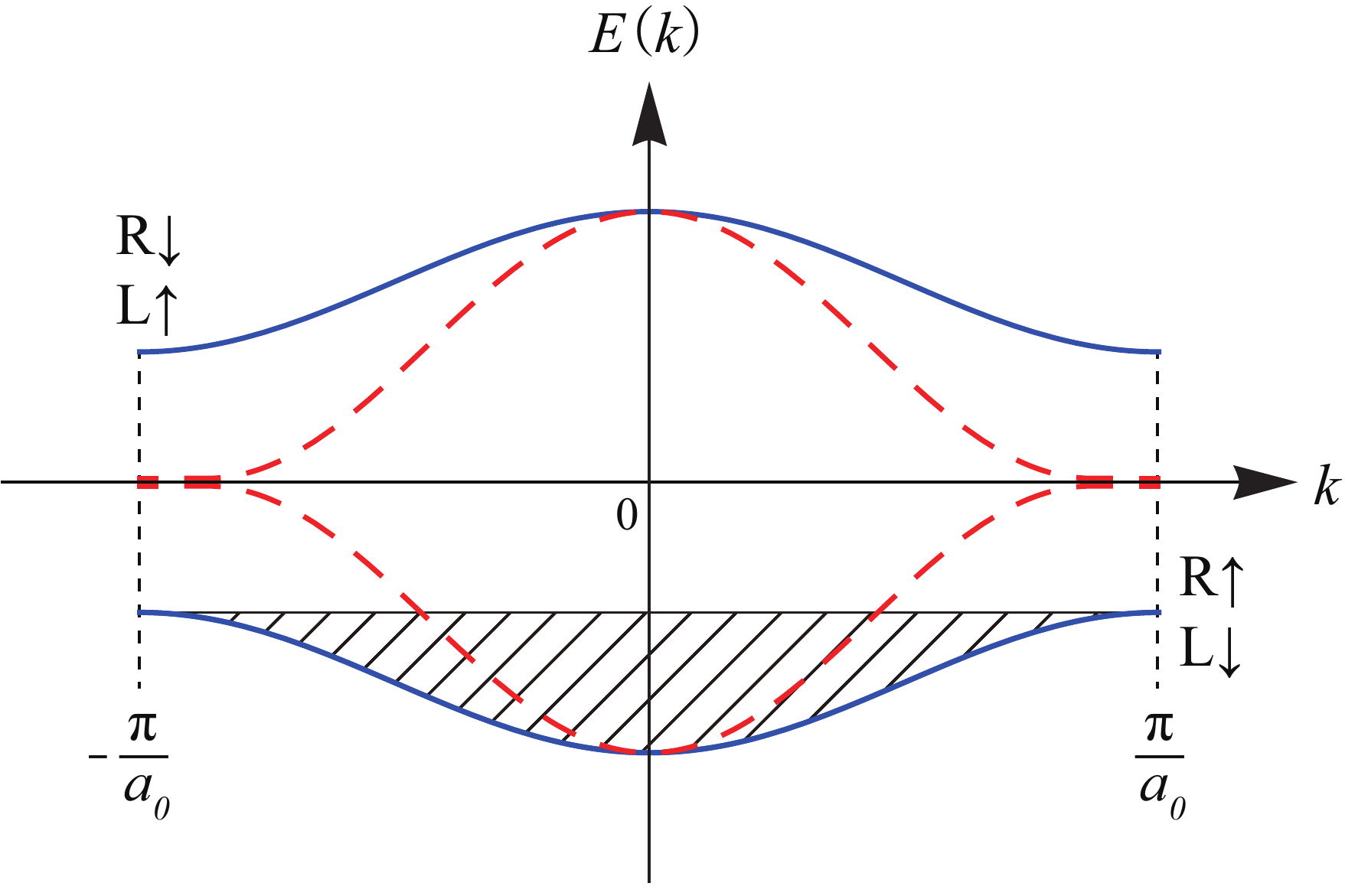}
\caption{Schematic illustration of band structure of interacting ZGNR in the absence of disorder. The valence (conduction) band is occupied (unoccupied). The gap size  is exaggerated and only bands near the gap are displayed. The natures of the unoccupied and occupied    states near $k=\pm \frac{\pi}{a_0}$  are given  ($a_0$   is the unit cell length of
the ZGNR), where $R$ and $L$ indicate states localized on the right and left zigzag edges, respectively. The arrows indicate spins, and the dashed lines represent the non-interacting bandstructure. }
\label{Ubandst1}
\end{center}
\end{figure}

\begin {figure}[!hbpt]
\begin{center}
\includegraphics[width=0.3\textwidth]{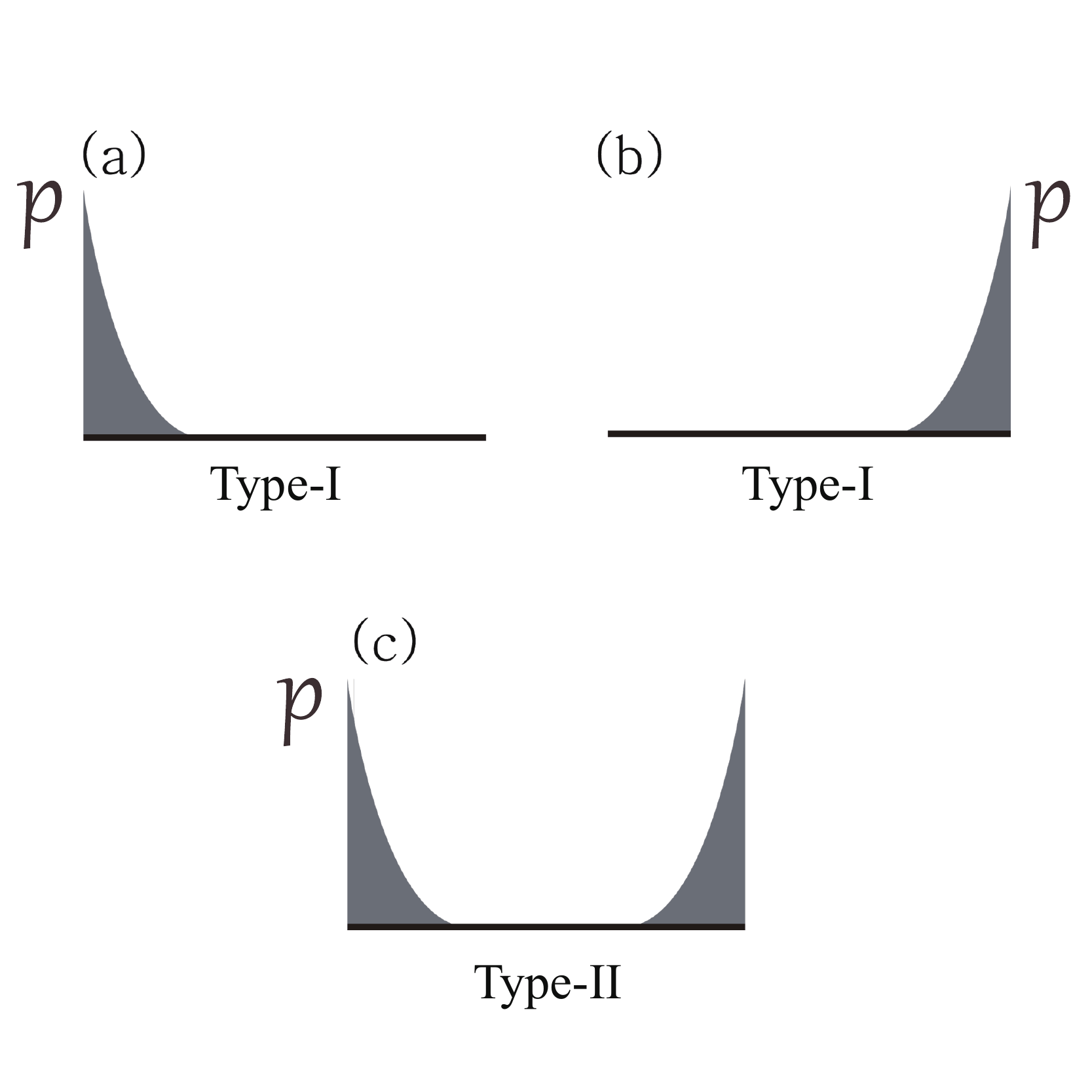}
\caption{   Schematic illustrations of site probability distributions $p$ of two degenerate edge states with wave functions $\phi_L$ and $\phi_R$,  localized on the (a) left and (b) right edges, respectively. In the absence of electron-electron interactions and disorder, there is some ambiguity in the selection of the  degenerate zigzag edge states.  (c) Another degenerate representation can be formed, with antibonding  $\frac{1}{\sqrt{2}}(\phi_L-\phi_R) $  and bonding   $\frac{1}{\sqrt{2}}(\phi_L+\phi_R)$ edge states.}
\label{degGap}
\end{center}
\end{figure}

In the following, we classify the possible zigzag edge states of  rectangular GNRs and periodic ZGNRs.
In the absence of electron-electron interactions and disorder, zigzag edge states near the Brillouin zone boundary  are   degenerate\cite{Brey,Neto} (see Fig.~\ref{Ubandst1}). For a given degenerate pair, there is some ambiguity in choice of single-electron wave functions: suppose $\phi_L $ and $\phi_R$ are two such degenerate states located on the left and right edges, respectively, as shown in Figs.~\ref{degGap}(a) and (b)   (we will classify $\phi_L $ and $\phi_R$  as Type-I states). If these states are chiral, one has A-carbon chirality and the other B-carbon chirality. However, the bonding 
$\frac{1}{\sqrt{2}} (\phi_L +\phi_R )$ and antibonding  $\frac{1}{\sqrt{2}} (\phi_L -\phi_R )$ edge states are 
also degenerate with the same energy, as shown in Fig.~\ref{degGap}(c) (we will classify these states as Type-II states). The probability density of these edge states is fractionalized equally between the left and right zigzag edges\cite{Jeong0}. This is similar to the fractionalization occurring at the end points of a long, insulating, one-dimensional wire\cite{Kit,Jack1}.

In the presence of electron-electron interactions, but without disorder,   an excitation gap $\Delta$ separates the conduction and valence bands\cite{Yang} (see Fig.~\ref{Ubandst1}).  In this case, there are  no states inside the gap, but  chiral zigzag edge states are present near the gap edges.  For the occupied states near $E\approx -\Delta/2$, one of the chiral edge states  of $\phi_L$ and $\phi_R$  is occupied by a spin-up electron while the other state is occupied by a spin-down electron, as shown in Fig.~\ref{Ubandst1}  (for the unoccupied states near $E\approx \Delta/2$,  the opposite holds).   As $\phi_L$ and $\phi_R$ are spatially separated, there is no repulsive energy and, therefore, these states  have  degenerate energy.  This is the physical origin of edge antiferromagnetism\cite{Waka1,Son}.

\begin{figure}[!hbpt]
\begin{center}
\includegraphics[width=0.35\textwidth]{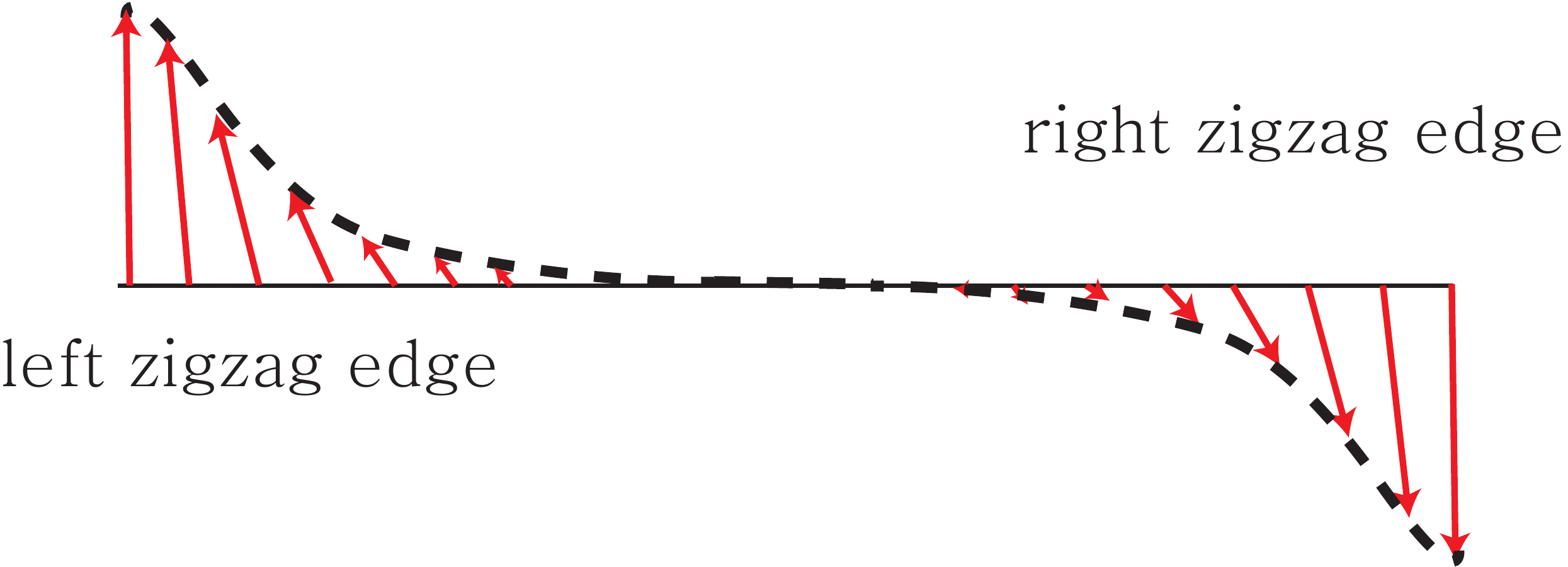}
\caption{Schematic drawing of site pseudospin probability rotation of Type-II edge state along  line connecting two left and right zigzag edges. The site pseudospin rotates out of plane by $\pi$ and behavior represents a kink.  It connects the left and right zigzag edges with different  magnetization directions,  analogous to a domain wall soliton in polyacetylene that connects two dimerized phases.}
\label{pseudo}
\end{center}
\end{figure}

We next consider the other case in which electrons move in a disorder potential in the absence of electron-electron interactions.  Impurities may also couple $\phi_L$ and $\phi_R$  and break the degeneracy between them.  The resulting edge states are approximately bonding and anti-bonding states\cite{Park}  (however, they are more localized along the GNR direction as a result of disorder). Such a state has opposite chiralities on the left and right zigzag edges and an approximately  half-integer edge charge.  This state is  classified as a Type-II state. Under certain conditions, it may    exhibit charge fractionalization of $1/2$\cite{Jeong0}  (a precise definition of fractional charge is given in Sec.~4).   
Disorder can also generate a third type of edge state for which the site probability distribution  is divided unevenly between two zigzag edges, i.e., a Type-III state. For Type-II and -III states, the variation of the site pseudospin\cite{pseudo} of each sublattice basis represents a topological kink\cite{def} (the left and right zigzag edges consist of
atoms from opposite sublattices).  As the coordinate position varies from one edge to the opposite edge, the wave-function chirality changes from that for A-carbon site to that for  B-carbon sites; i.e., the site pseudospin rotates by $\pi$, as  shown schematically in Fig.~\ref{pseudo}.

To investigate  the formation of a gap-edge state with a fractional edge charge of $1/2$, both electron-electron interactions and disorder must be incorporated in a self-consistent manner.  Such a model is described in the next section.

\section{Model}

We consider a periodic ZGNR with $N$ carbon atoms, length $L$ and width $w$. As mentioned above, electron-electron interactions play an important role in ZGNRs as they induce an energy gap $\Delta$\cite{Waka1,Yang}. The system becomes an insulator in the half-filled case. Electron-electron interactions are modeled through the on-site repulsion $U$, the value of which depends on the type of substrate on which the GNR is formed (a substrate can provide screening of 
electron-electron interactions).  The zigzag edge magnetism is larger for stronger   $U$ values.

It is well-known that the range  of the impurity potential is important for determining the properties of the Dirac electrons in ZGNRs\cite{Ando,Lima,LongShort}.  Here, we model disorder  by placing  $N_{imp}$ defects or impurities randomly at carbon 
sites $\vec{R}_j$. We take the following simple  discrete model for the disorder potential 
at lattice site $\vec{r}_i$:
\begin{equation}
V_i=\sum_j^{N_{imp}}\epsilon_j e^{-|\vec{r}_i-\vec{R}_j|^2/d^2},
\label{dispot}
\end{equation}
where $d$ is the range of the   potential.    Note that, when  $d=0$,  the disorder potential   is defined  such that it is finite only    for $\vec{r}_i=\vec{R}_j$ (in a continuous model,  this  implies that the effective range is equal to the size of a carbon atom $\sim a_0$).  When $d\sim a_0 (d\gg a_0)$, the potential is short-(long)-ranged (here,
$a_0=1.73a$   is the unit cell length of
the ZGNR and $a=1.42$ \AA\  is the C-C distance).  The strength
of the  potential $\epsilon_j $ is chosen randomly from the energy interval $[-\Gamma,\Gamma]$.  The  values of $\epsilon_j $ and $d$ depend on the type of  charged impurity in the substrate and the defects in the graphene.  The defects have  $d\sim a_0$ while the impurities have  $d\gg a_0$.   In the self-consistent Born approximation, the disorder strength is characterized by the parameter $\Gamma\sqrt{n_{imp}}$\cite{SCBA}, where $n_{imp}=N_{imp}/N$. $N_{imp}$ is also relevant for determining the strength of the disorder potential.  Defects with a short-ranged potential are more relevant to the quantization of a fractional charge, as demonstrated below.

We include both electron-electron interactions and disorder in a tight-binding model at half-filling.
 The interplay between  $U$ and the disorder can be  treated using the self-consistent Hartree-Fock approximation (HFA).  When $U=0$, the   disorder can be  treated exactly  by this method. When the disorder is absent, the interaction effects can be well represented  by the HFA, which is widely used  in graphene-related systems\cite{Ross,Pis,Sor,Stau}.  The results are consistent with those of 
density functional theory\cite{Yang}.   When both disorder and interactions are present, the self-consistency provides an excellent approximation\cite{Eric}.  
The total Hamiltonian in the HFA is
\begin{eqnarray}
&&H=-t\sum_{<ij>\sigma} c_{i\sigma}^{\dag}c_{j\sigma}+\sum_{i\sigma} V_ic_{i\sigma}^{\dag}c_{i\sigma}\nonumber\\
&+&U\sum_{i\sigma}  (n_{i\uparrow} \langle n_{i\downarrow}\rangle+\langle n_{i\uparrow}\rangle n_{i\downarrow}
-\langle n_{i\uparrow}\rangle \langle n_{i\downarrow}\rangle )\nonumber\\
&-&\frac{U}{2}\sum_i (n_{i\uparrow} +n_{i\downarrow} ),
\label{Ham}
\end{eqnarray}
where  $c_{i\sigma}^{\dag}$ and $n_{i\sigma}$ are the electron creation and  occupation operators, respectively, at site $i$ with spin $\sigma$. As the translational symmetry is broken, the Hamiltonian is written in terms of the site representation.  In the hopping term, the summation is over the nearest neighbor sites (the hopping parameter is $t\sim 3\textrm{ eV}$).
The eigenstates and eigenenergies  are computed numerically by  solving the tight-binding Hamiltonian matrix self-consistently.  
The self-consistent occupation numbers $\langle n_{i\sigma}\rangle$ in the Hamiltonian are
the sum of the probabilities of finding electrons of spin $\sigma$ at site $i$:
\begin{eqnarray}
\langle n_{i\sigma}\rangle=\sum_{E_{\sigma}\leq E_F} |C_i (E_{\sigma}) |^2 .
\label{occnum}
\end{eqnarray}
The performed sum is over  the energy of the occupied eigenstates  $E_{\sigma}$ below the Fermi energy $E_F$. 
Note that  $\{C_i(E_{\sigma})\}$ 
represents an eigenvector of the tight-binding Hamiltonian matrix with  $E_{\sigma}$.   For notational simplicity, we suppress  its dependence on $E_{\sigma}$ here after.    We  define 
the weak disorder regime as that in which the
ratio between the disorder strength and interaction strength is
$\kappa\equiv\Gamma \sqrt{n_{imp}}/{U}\ll 1$.
In this work, the band is half-filled: the filling factor is $f=N_e/2N=1/2$ (where $N_e$ is the number of electrons).   Below, we use $U/t=1$ unless stated otherwise.  

\section{Quantized fractional charge}

Using the eigenvectors $\{C_i\}$, we  define the fractional charge in the {\it many-body ground state}.  If an electron is added into or removed from a gap state, it can become fractionalized into charges located on the left and right zigzag edges. Suppose $\langle n_i^a\rangle$ is the  ground state site occupation number {\it after} an electron with spin $\sigma$ and  energy $E$ is removed from a gap state represented by  an eigenvector $\{C_i\}$.   It can be written as
\begin{eqnarray}
\langle n_i^a \rangle=\langle n_i \rangle-|C_i  |^2,
\end{eqnarray}
where $\langle n_i\rangle=\sum_{\sigma}\langle n_{i\sigma}\rangle$ is the ground state site occupation number {\it before} the removal.
The fractional boundary charge  is defined using this $\langle n_i^a \rangle$\cite{Kiv}, such that
\begin{eqnarray}
q=\sum_i f_i (1-\langle n_i^a \rangle).
\end{eqnarray}
Here, the average site occupation is set to $1$. For the fractional charge located on the left edge, the sampling function $f_i $ is centered around the fractional charge. 
The fractional charge located on the right edge is similarly defined. The left and right fractional charges decay exponentially, and their overlap is negligible for  large  $w$. We can rewrite $q$ as
\begin{eqnarray}
q=\sum_i f_i (1-\langle n_i\rangle) + \sum_i f_i |C_i|^2 = \sum_i f_i |C_i|^2.
\end{eqnarray}
Here, as a result of  the random potential, the sum $\sum_i f_i(1-\langle n_i\rangle)$ vanishes because the site occupation number fluctuates around the mean value: $\langle n_i\rangle = \langle n_{i\uparrow}\rangle +\langle n_{i\downarrow}\rangle = 1\pm\delta_i$, where $\delta_i$ is a random number. The resulting fractional charge is equal to the total probability on the A- or
B-carbon sites:
\begin{eqnarray}
q_A=\sum_{i\in A} |C_i|^2,  \quad
q_B=\sum_{i\in B} |C_i|^2.
\label{qch}
\end{eqnarray}
Note that $q_{A,B}$ depend on  $E_{\sigma}$. For Type-I, either $q_A\approx 1$ or $q_B\approx 1$, and for Type-II,  $q_A\approx 1/2$ and $q_B\approx 1/2$.  For Type-III state, $q_A$ can deviate significantly from $1/2$ and $1$.  For any type  of  gap state, $q_A + q_B = 1$, which follows from the normalization of the wave functions.  In the following, we  denote disorder-averaged values as $\overline{q}_A$  and $\overline{q}_B$.

It is important that the fractional charge is quantized to a high precision\cite{Kiv}.   Several conditions must be met: 1) the fractional charges must not overlap, 2) the fractional charge profile must decays rapidly, and 3) the charge fluctuation  must be small. Note that, the wider the ZGNR, the better the fractional quantization, as the overlap between the fractional charges on the left and right zigzag edges decreases.  In addition,  the edge wave functions decay 
exponentially\cite{Neto}, as indicated by their site probability distributions (see below).  The charge fluctuations are estimated in Sec.~6.

\section{Properties of  gap-edge states }

\begin{figure}[!hbpt]
\begin{center}
\includegraphics[width=0.2\textwidth]{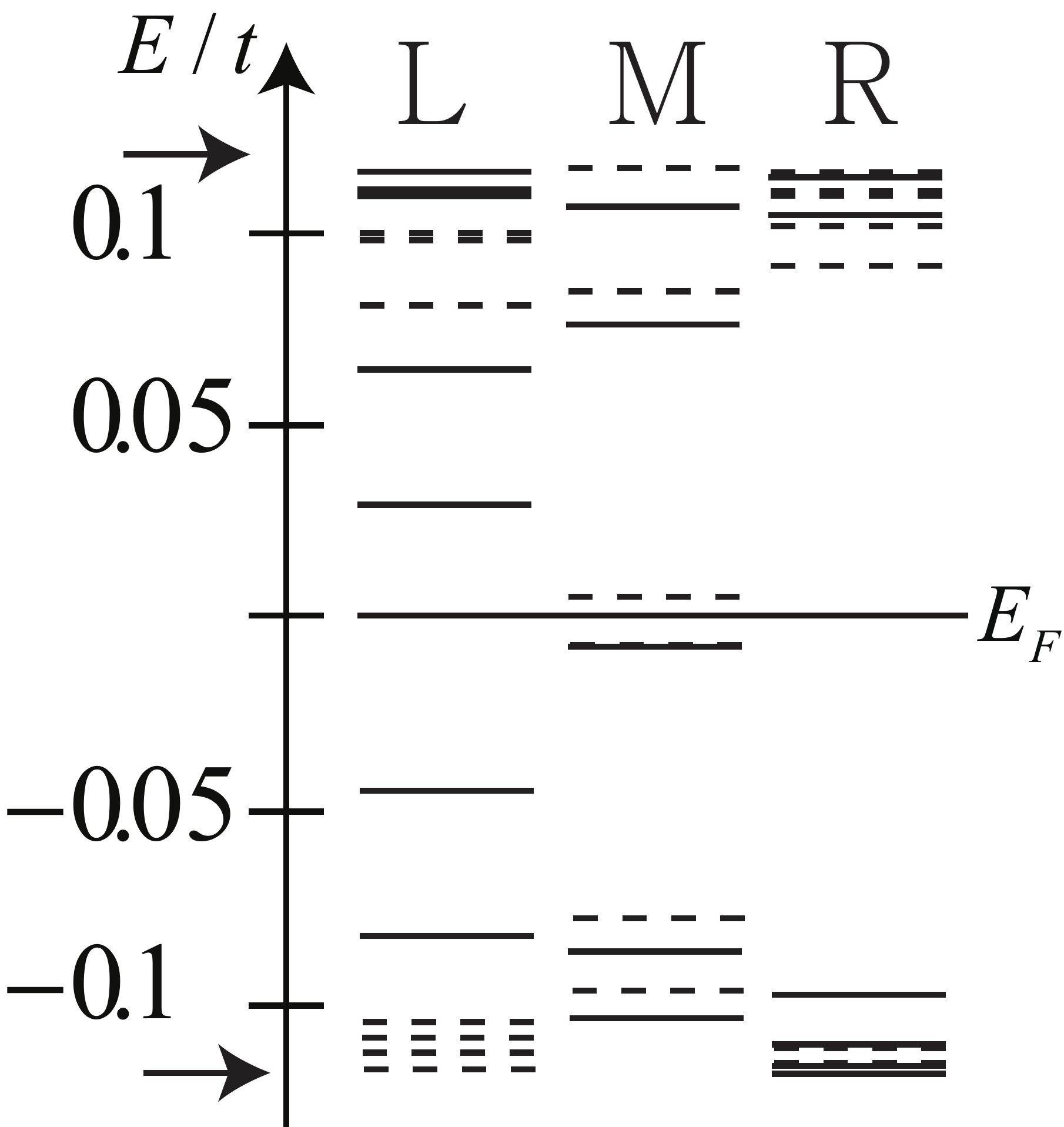}
\caption{ 
Gap states in interval $|E|/t \lesssim \frac{\Delta}{2t}=0.105$
for $\Gamma/t=0.35$ (in the absence of disorder,  $E_F=0$).  The horizontal arrows indicate the gap edges and the solid (dashed) lines represent the spin-up (-down) levels. Here $L=125.4$ \AA,   $d=0$, $n_{imp}=0.1$, and $w=7.1$ \AA.  The disorder realization number is $N_D=1$.  L (R) represents  Type-I and -III gap-states  dominantly localized on the left (right) zigzag edge. M represents  Type-II gap-edge states  divided between the left and right zigzag edges. The straight (dashed) lines indicate the spin-up (-down) states. Note that the thick lines represent degenerate states.}
\label{spec}
\end{center}
\end{figure}

\begin{figure}[!hbpt]
\begin{center}
\includegraphics[width=0.35\textwidth]{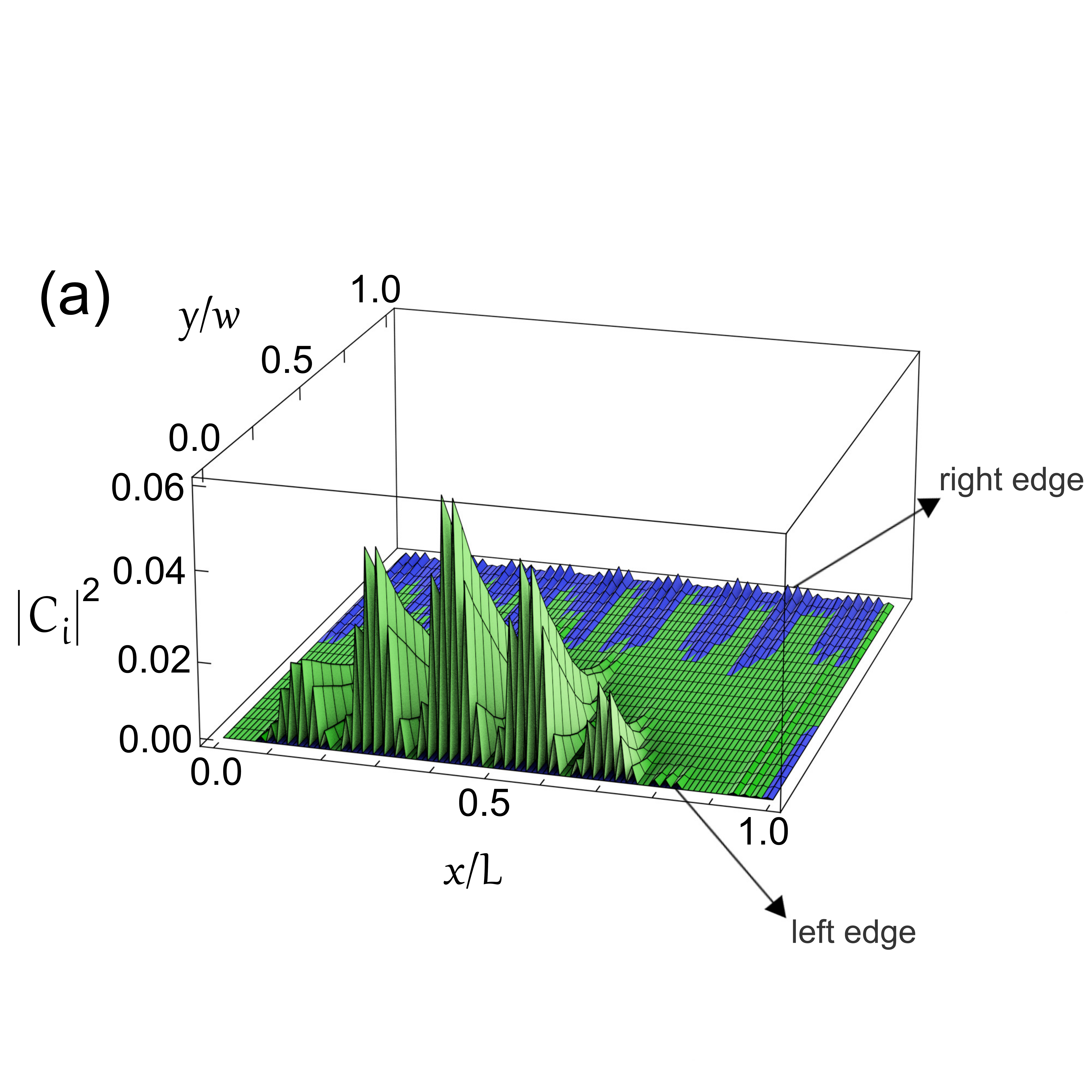}
\includegraphics[width=0.35\textwidth]{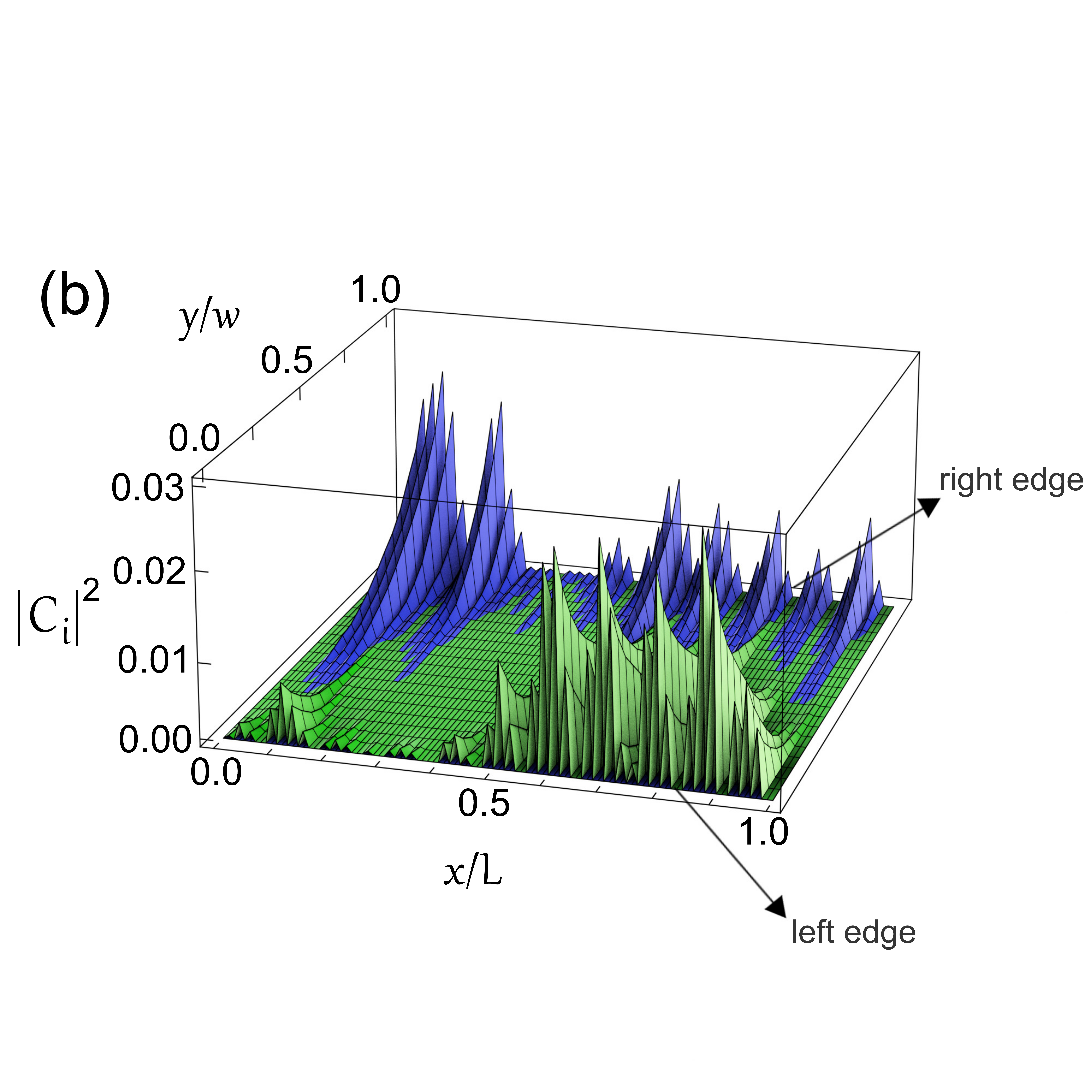}
\caption{(a) Site probability distribution of Type-I edge state near gap edge  with energy $E/t=-0.107$  for $\Gamma/t=0.5$, $d=0$, and $n_{imp}=0.01$.  The green (blue) color represents the probability on the A(B)-carbon sites.
(b) Site probability distribution   of Type-II edge state outside  gap  for $\Gamma/t=0.5$, $d=0$, and  $n_{imp}=0.1$. Here, $E/t=-0.154$.    In both figures,
$L=125.4$ \AA, and $w=7.1$ \AA. }
\label{outgapstate}
\end{center}
\end{figure}

\begin{figure}[!hbpt]
\begin{center}
\includegraphics[width=0.35\textwidth]{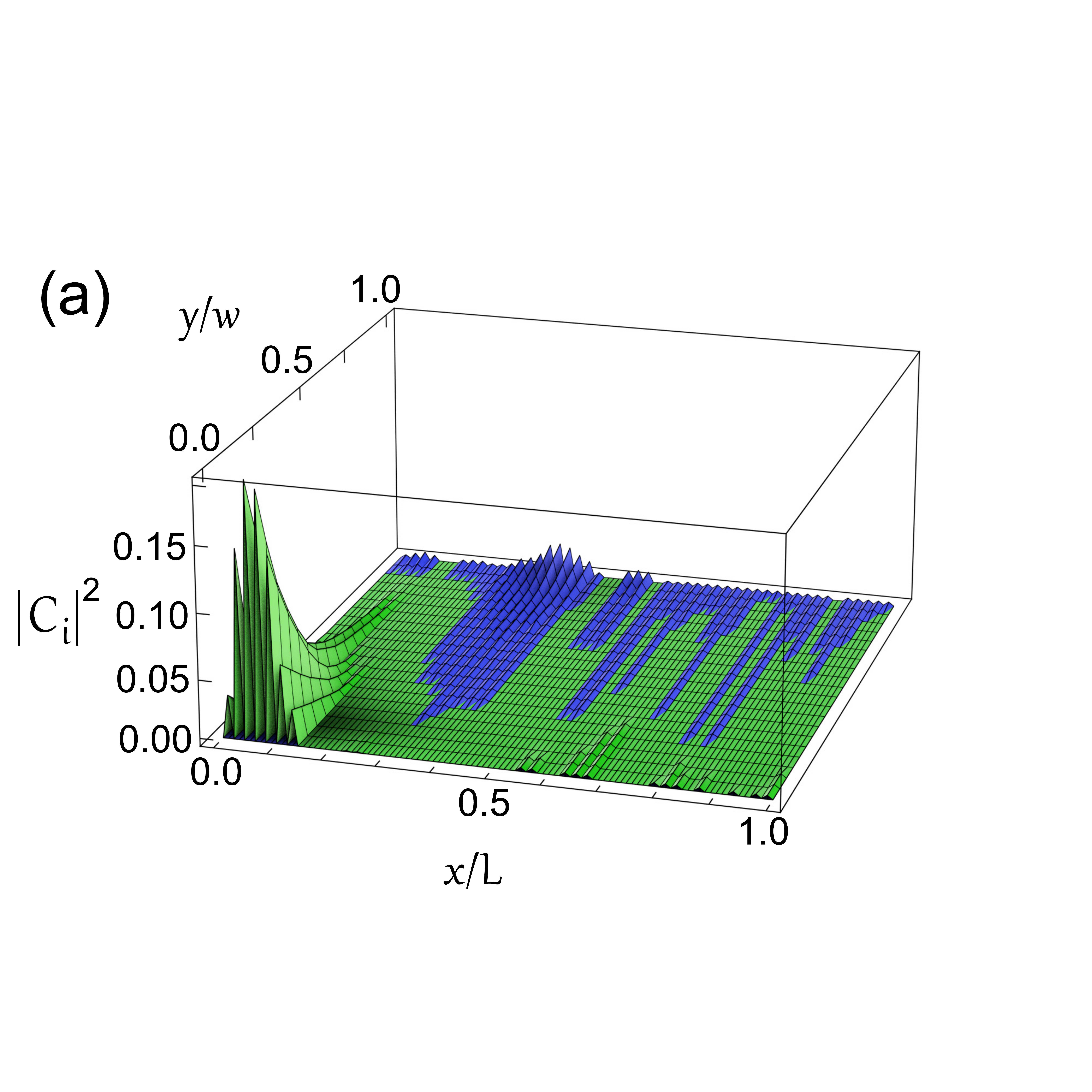}
\includegraphics[width=0.35\textwidth]{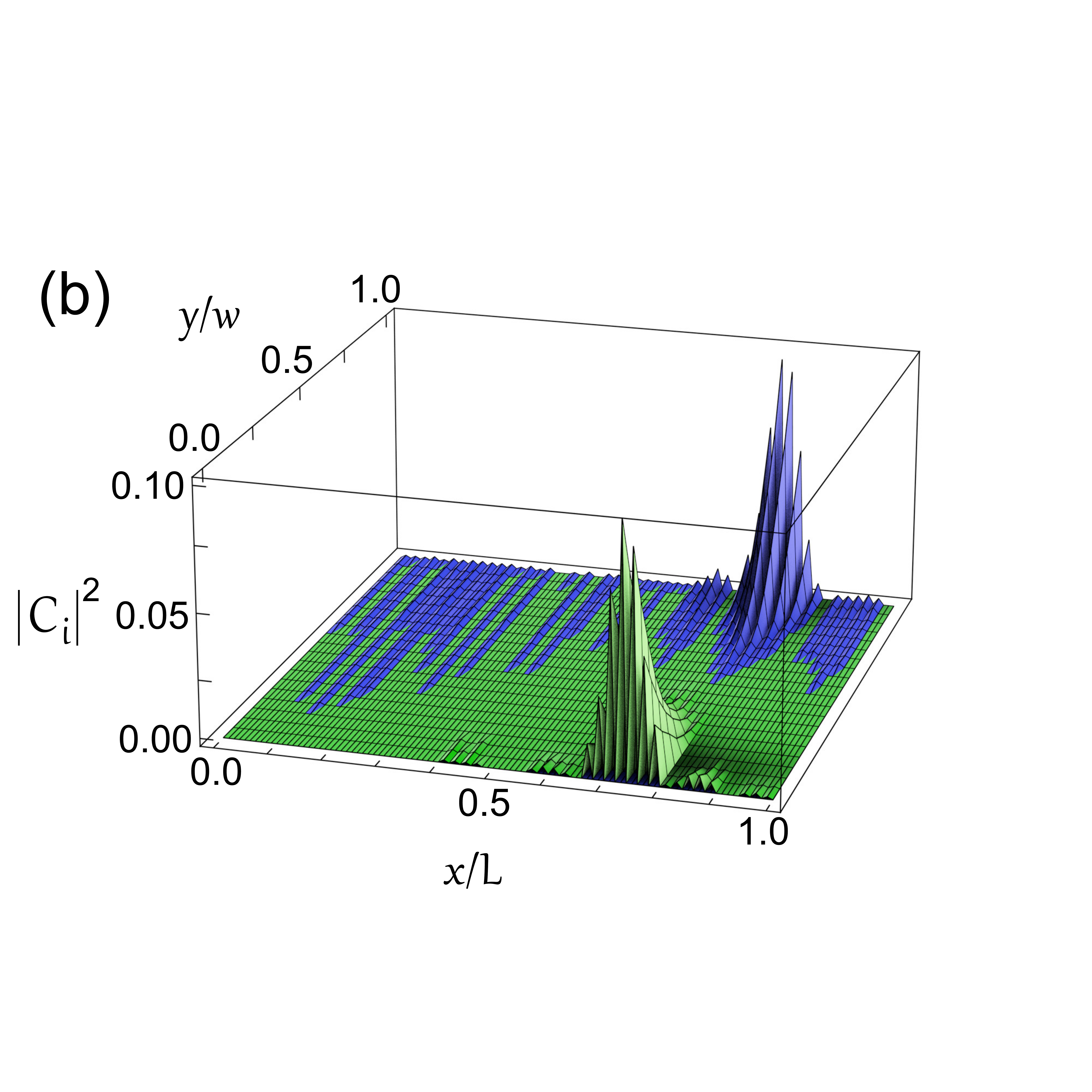}
\caption{Site probability distributions   of localized (a) Type-I  and  (b) -II gap-edge states with  $E/t=-0.071$  and $0.024$, respectively. The parameters are $d=0$,
$n_{imp}=0.01$, $\Gamma/t=2.0$, $L=125.4$ \AA, and $w=7.1$ \AA.  
}
\label{ingapstate}
\end{center}
\end{figure}

To study the  formation of a fractional charge, we must plot the site probability of the   $\{C_i\}$ of the gap-edge states 
(the site probability distribution is given by  $|C_i|^2$).   The shape of the site probability distribution indicates the number of  charge fractions located on the zigzag edges.  For this purpose, we must analyze the  gap-edge states  of a single disorder realization (their energy spectrum is given in Fig.~\ref{spec}). 

The site probability distributions  of two states near the gap energy $|E|\approx \Delta/2$ are plotted in Fig.~\ref{outgapstate}.
That shown in Fig.~\ref{outgapstate}(a) is  for  $n_{imp}=0.01$ and $\Gamma=0.5t$. The site probability decays along the perpendicular direction to the left of the zigzag edge.  This state is chiral with significant probability on the A-carbon sites only (Type-I).
The edge state  shown in Fig.~\ref{outgapstate}(b) is in the intermediate   disorder regime with  $n_{imp}=0.1$ and $\Gamma=0.5t$. The  site probability  is significant on both the left and right zigzag edges.  However, the overlap is not negligible.  Note that opposite chiralities are obtained on the left and right zigzag edges.

\begin{figure}[!hbpt]
\begin{center}
\includegraphics[width=0.35\textwidth]{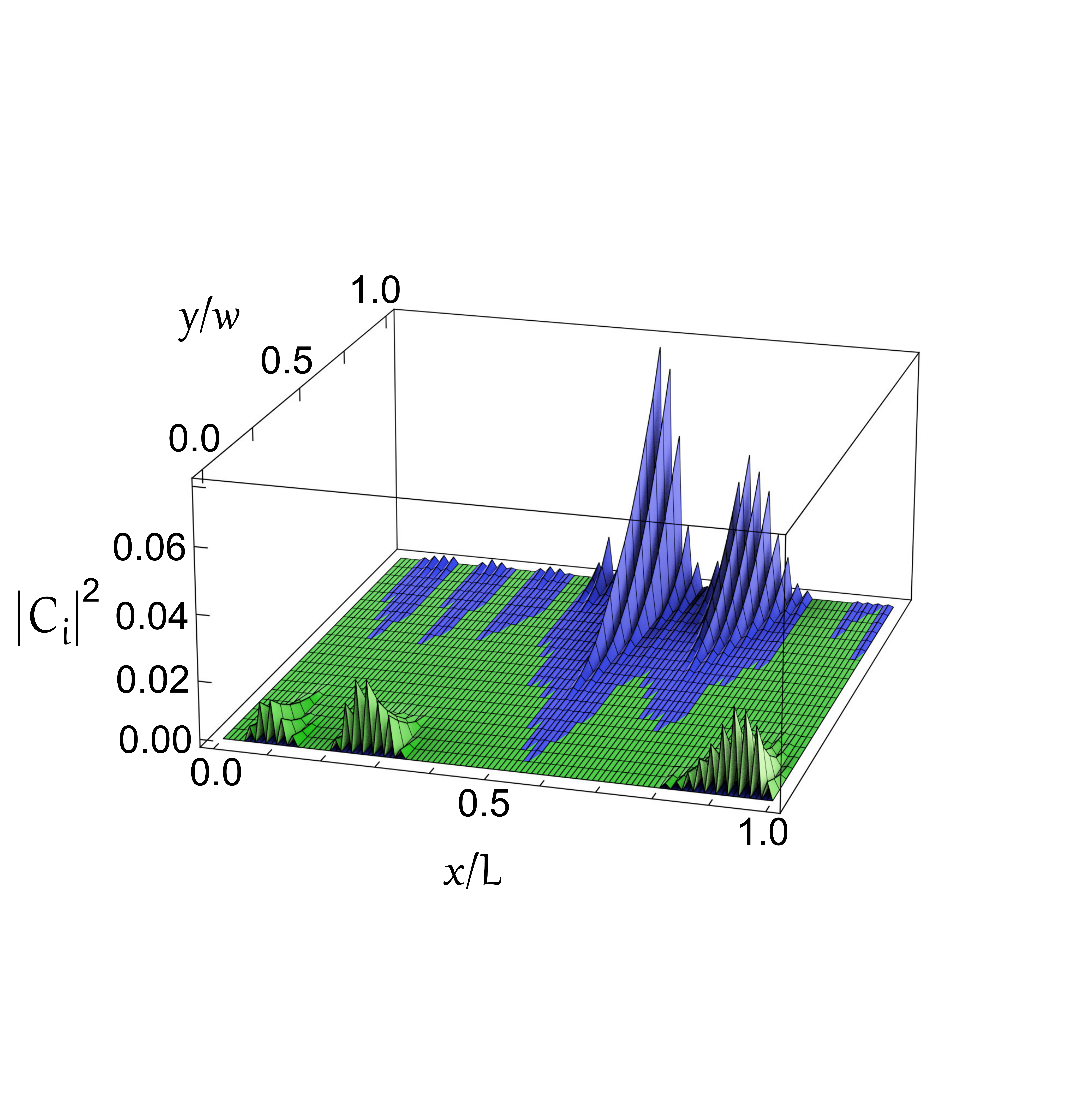}
\caption{Site probability distribution   of  localized Type-III gap-edge state with  $E/t = 0.11$. The parameters are $d = 0, n_{imp} = 0.1, \Gamma/t = 0.5, L = 125.4$ \AA, and $w = 7.1$ \AA.}
\label{type3st}
\end{center}
\end{figure}

\begin{figure}[!hbpt]
\begin{center}
\includegraphics[width=0.3\textwidth]{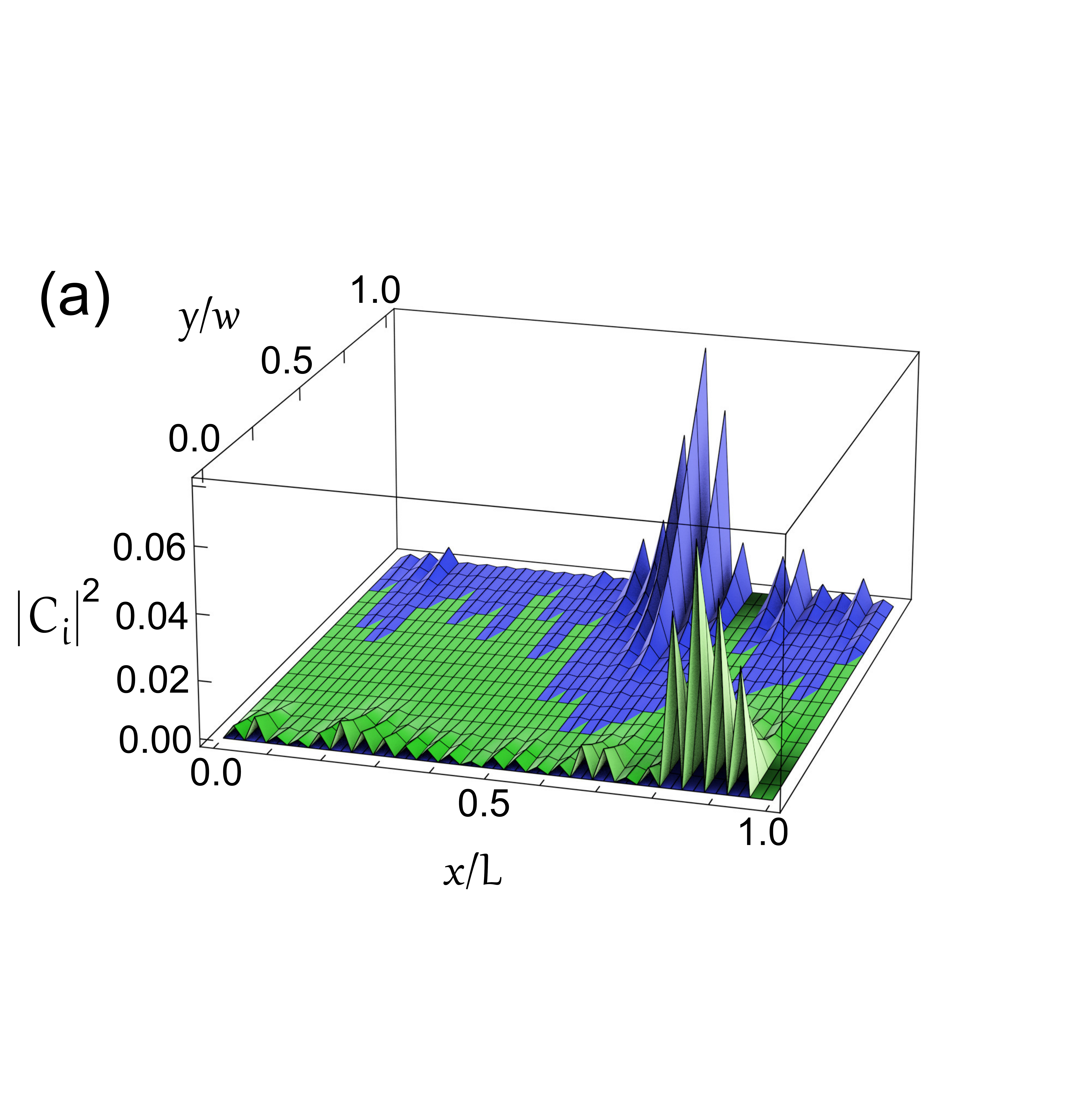}
\includegraphics[width=0.3\textwidth]{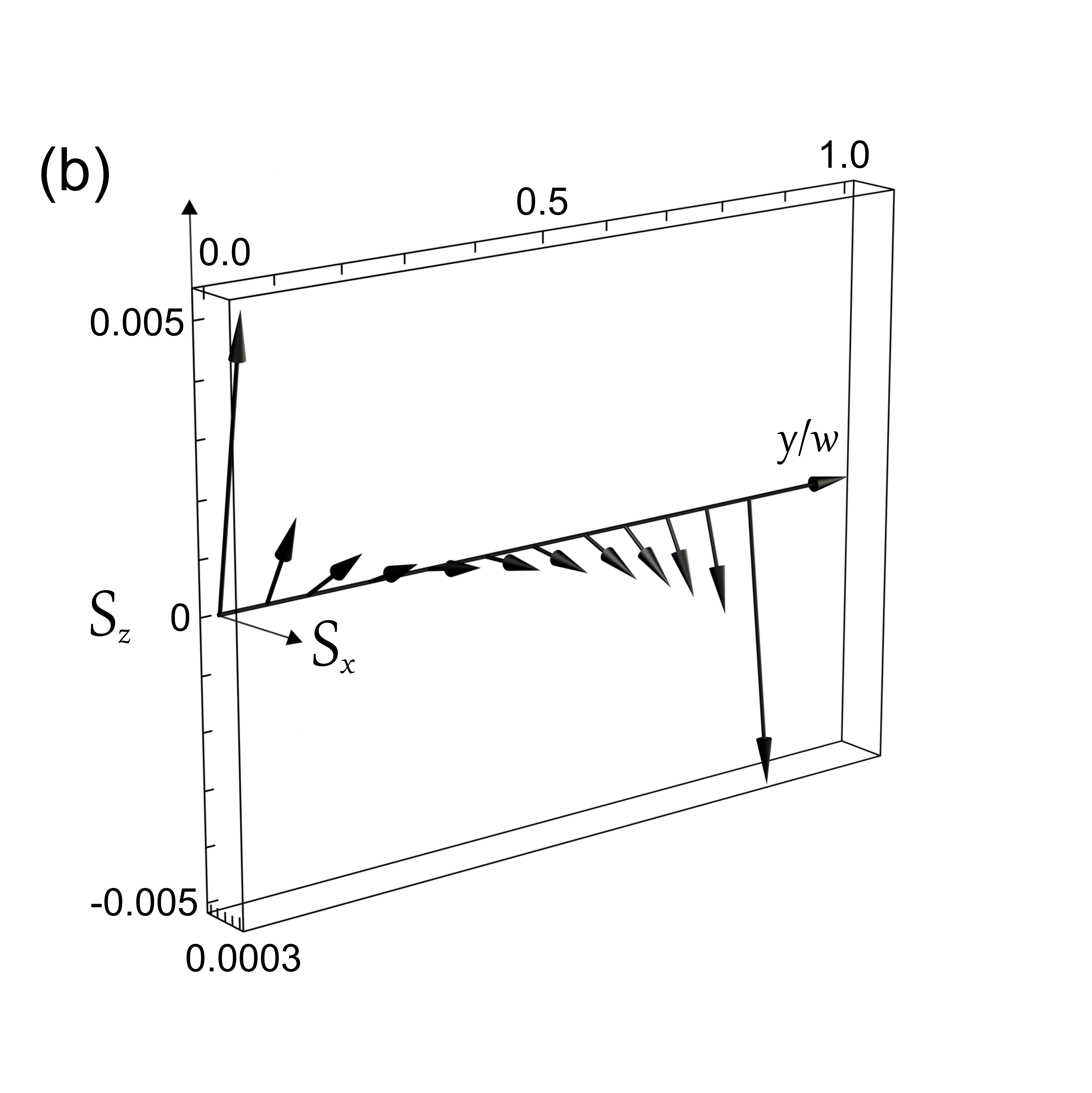}
\caption{ (a) Site probability distribution   of Type-II gap-edge state.   The parameters are $L=59.46$ \AA,  $w=24.14$ \AA, $d=0$, $n_{imp}=0.1$,  and  $\Gamma/t=0.5$.
(b) Site pseudospin values $(S_x,S_y,S_z)$ of this Type-II state at sublattice basis positions along  y-axis. The values are averaged along the x-axis.  Note  that
$S_y=0$. }
\label{pseu}
\end{center}
\end{figure}

\begin{figure}[!hbpt]
\begin{center}
\includegraphics[width=0.3\textwidth]{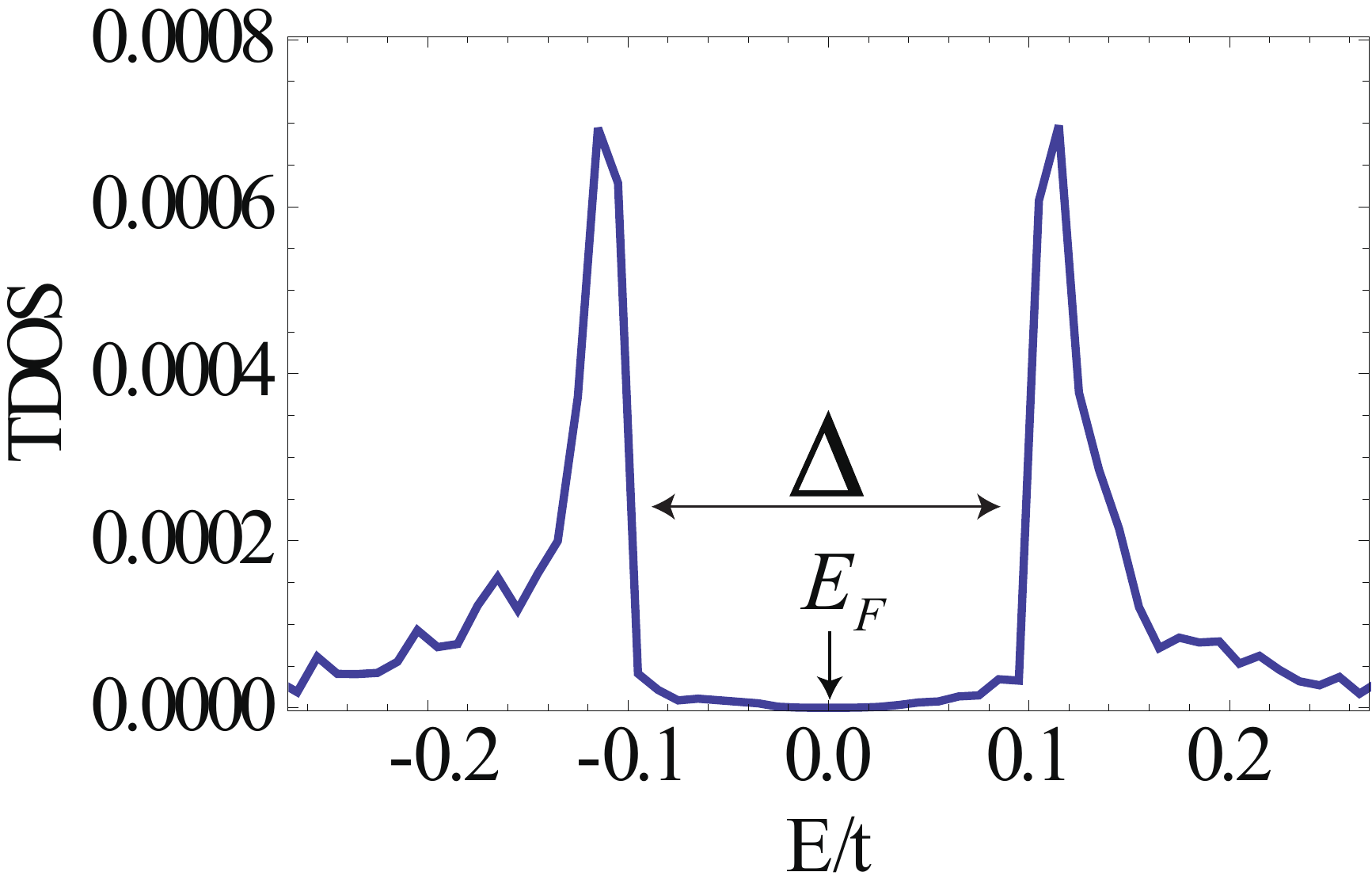}
\caption{Typical density of states (TDOS) for disorder strength $\Gamma=0.5t$,  with  $N_{D}=800$,    $n_{imp}=0.01$, and   $d=0$. For the ribbon, $w=7.1$ \AA, and $L=307.4$ \AA.  The  histogram energy interval is $\Delta E/t=0.01$.}
\label{tdos}
\end{center}
\end{figure}

Edge states also exist within the  gap, i.e., with  $|E|< \Delta/2$. The site probability distributions   of two gap-edge states in the intermediate disorder regime with $n_{imp}=0.01$  and $\Gamma=2t$ are plotted in Fig.~\ref{ingapstate};  these are both  localized.  The state shown in Fig.~\ref{ingapstate}(a) is localized without fractionalization (Type-I).
The site probability distribution  of the  state shown in Fig.~\ref{ingapstate}(b) is approximately  fractionalized into two parts, each with an approximate value of $1/2$ near the left and right zigzag edges (Type-II).  Note also that their overlap is negligible (the overlap between the fractional charges on the left and right zigzag edges decreases even further when the ZGNR width is larger)
A {\it charge fractionalization} of $1/2$ is thus expected for this {\it midgap} state\cite{Kit}.

Another edge state (Type-III) is displayed in Fig.~\ref{type3st}.  Its site probability  is distributed unevenly between the left and right zigzag edges.  Note that, when the disorder potential breaks the inversion symmetry, analysis of the Zak phase suggests that the boundary charge may deviate from integer and half-integer values\cite{Zak,Jeong1,Van}.   

Type-II and -III gap-edge states  correspond to a topological kink.  The site pseudospin value\cite{pseudo}  rotates by approximately $\pi$ as the coordinate  position varies from one edge to the opposite edge\cite{Jeong0}; therefore, the chirality of the wave function changes from that of the A-carbon sites to that of the  B-carbon sites.   An example  is shown  in Fig.~\ref{pseu}.  In interacting disordered ZGNRs, Type-II and -III kinks  coexist.  A kink is rather similar to a soliton, which corresponds to a domain wall connecting two different dimerized phases of  polyacetylene\cite{Sol}.

A short-ranged disorder potential affects the  localization properties and changes  the site probability distribution along the ribbon direction\cite{Lima}.  The  localization properties can be studied by computing the  disorder-averaged value of  
the typical DOS (TDOS)\cite{Jan}.  A finite TDOS in the limit $L \rightarrow \infty$ is an indication of delocalized states.
The TDOS   for $L = 307.4$ \AA\ is shown as a function of $E$ in Fig.~\ref{tdos}. The TDOS is nearly zero in the interval $|E|\lesssim\Delta/2$ and displays little finite-size dependence on $L$, which suggests that the corresponding states are localized. The TDOS changes rapidly near  $E\approx \pm\Delta/2$, suggesting that the critical energy of the localization/delocalization transition is $E_c\approx\Delta/2$ (in the absence of disorder, the states with   $|E|>\Delta/2$  are delocalized).  Note that the ordinary one-dimensional localization theory\cite{Abr} does not apply here\cite{com}.

It is interesting to note  that most of the gap states are {\it spin-split}, i.e.,  
singly occupied, as apparent from  Fig.\ref{spec} (a Mott-Anderson insulator  also possesses singly occupied states\cite{Dob}). Some states are nearly spin degenerate, but their wave functions are different.
In GNRs, it appears that the gap states are spin-split when {\it both} zigzag edges and an external potential are present\cite{Jeong1, Sor}.   Conversely, the states far  from the gap (non-edge states) are almost spin degenerate.   Spin splitting may affect edge antiferromagnetism.  The magnetic properties of the gap states are given in Appendix.

\section{Edge charge distribution}

We next perform  disorder averaging over numerous disorder realizations to obtain  the distributions of Type-I, -II, and -III states as functions of eigenenergy.  We also compute the number of  gap states having    edge charge values $q_A$ or $q_B$, defined in Eq.~(\ref{qch}).   The gap-edge states that are localized on the left zigzag edge have $1/2<q_A<1$ and $0<q_B<1/2$. (Conversely, the edge states localized on the right edge have $0<q_A<1/2$ and $1/2<q_B<1$.)  
These results may show the eigenenergy  at which  fractional charge quantization may appear.

\begin{figure}[!hbpt]
\begin{center}
\includegraphics[width=0.3\textwidth]{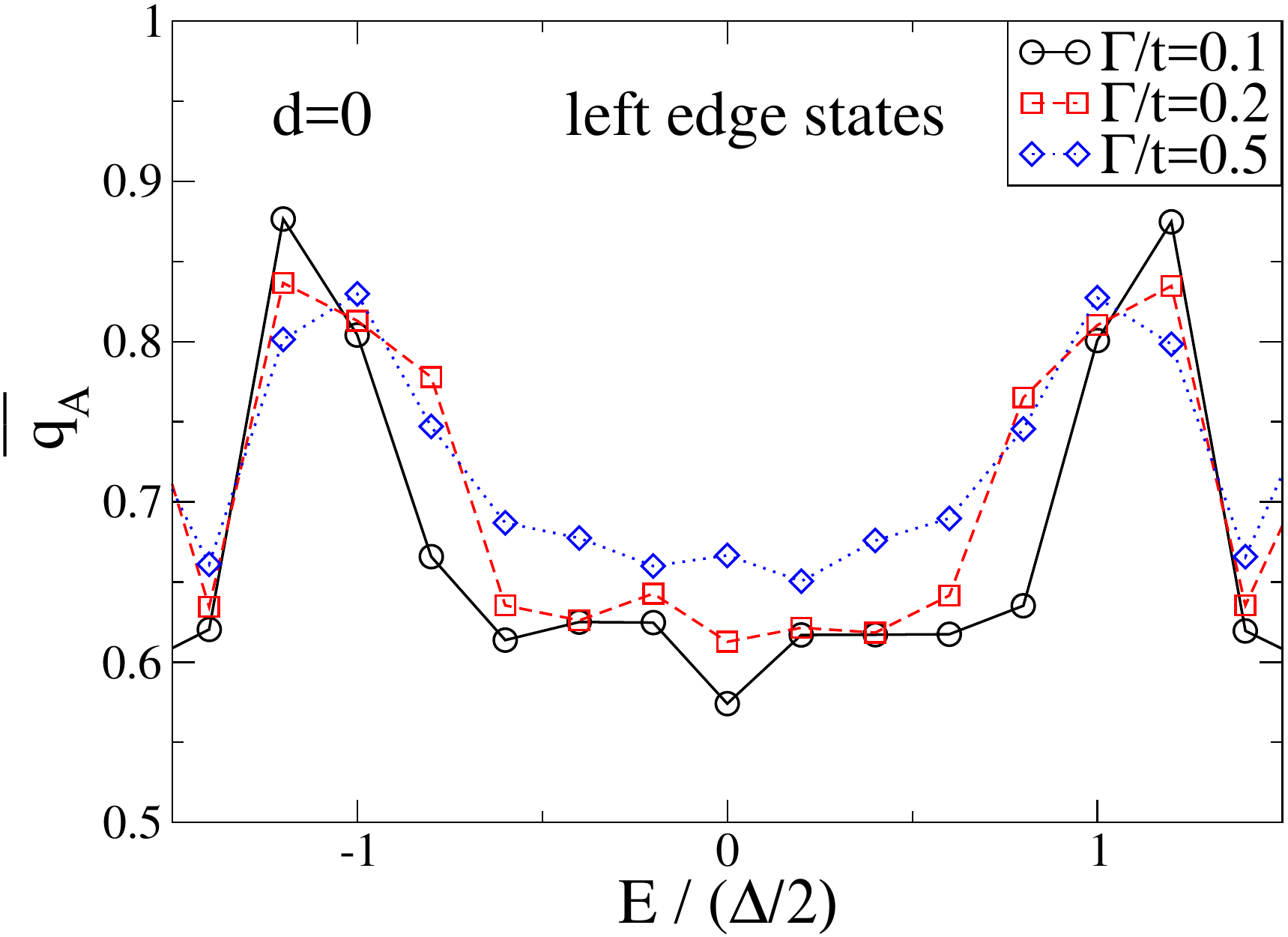}
\caption{ Disorder-averaged value of $\overline{q}_A$  of  {\it left} edge states for $d=0$. The  parameters are $N_D=1000$, $n_{imp}=0.1$, $w=7.1$ \AA, and $L=125.4$ \AA.   The histogram interval is $\delta E=0.1\Delta$.    Note that some states with energy outside the energy gap are also included.}
\label{qA0}
\end{center}
\end{figure}

First, let us compute the disorder-averaged value 
$\overline{q}_A$ as a function of $E$.  We count the left zigzag edge states in the energy interval $[E-\frac{\delta E}{2},E+\frac{\delta E}{2}]$  and compute the average value  $\overline{q}_A$.  Fig.~\ref{qA0} displays  $\overline{q}_A$ as a function of $E$ for $d=0$.  For small values of $\Gamma$, it is apparent that the  value  of
$\overline{q}_A$   takes the  minimum value for the midgap states  (the energies of midgap states  are in a small  energy interval $[-\frac{\delta E}{2},\frac{\delta E}{2}]$ around  $E_F=0$).
When {\it both} the left and right edge states are counted
in each energy interval,   $\overline{q}_A$  is close to $1/2$, {\it independent} of $E$.   For example, let us consider states near the gap edge $E=-\frac{\Delta }{2}$.  If one  is localized on the left zigzag edge with 
$q_A\approx 1\  (q_B\approx 0)$,  there is also a state that is localized on the right edge with 
$q_B\approx 1\ (q_A\approx 0)$.  The resulting mean value of $q_A$ is   then
$1/2$.  
Counting both the left and right edge states, we find that   the states in the middle of the gap are  mostly of  Type II, while the states  near the gap edges are mostly of Type I.  At other energies, the Type-III states are distributed broadly between $E=0$ and $\pm \Delta/2$.

\begin{figure}[!hbpt]
\begin{center}
\includegraphics[width=0.3\textwidth]{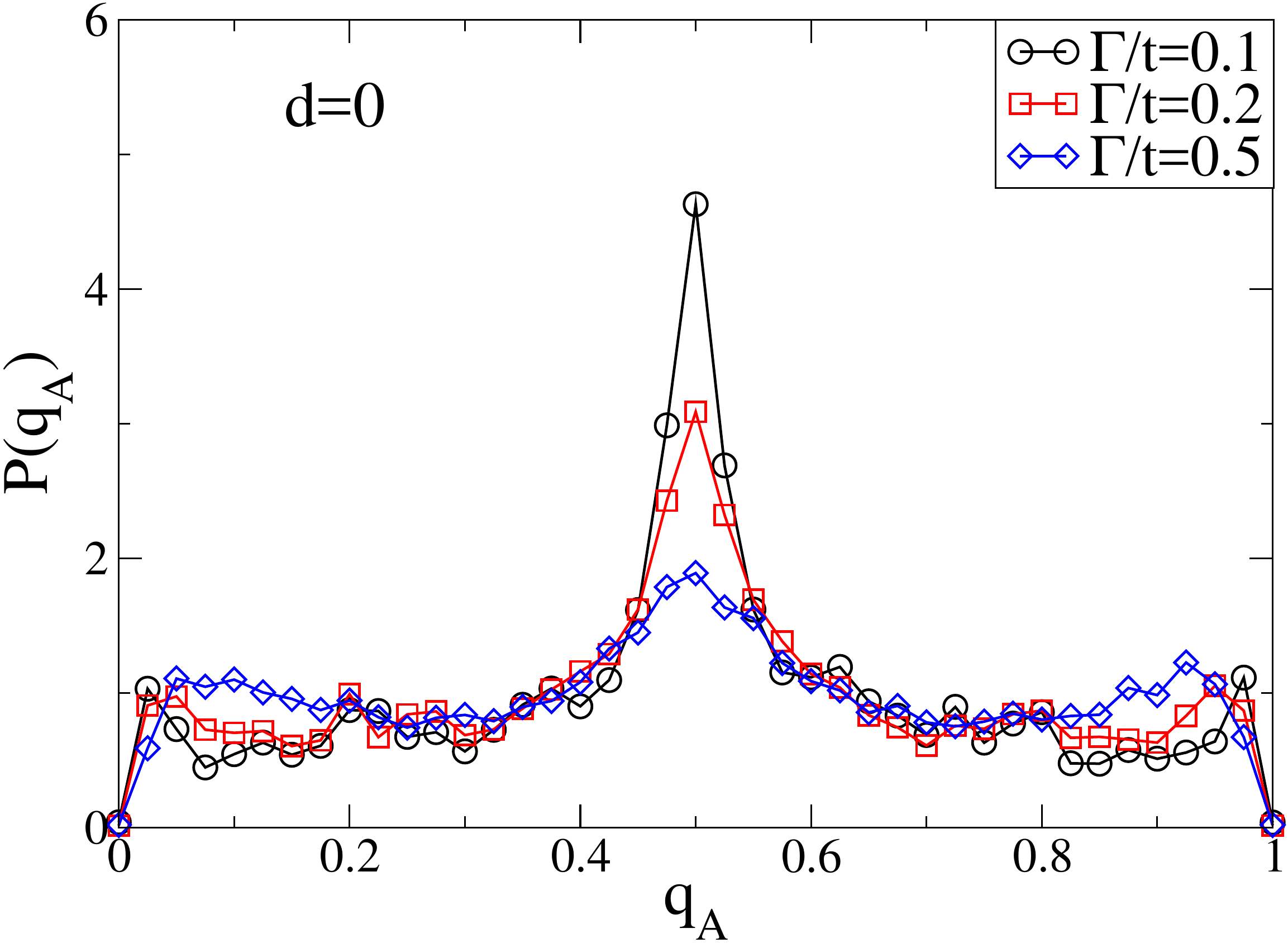}
\caption{Probability distribution of disorder-averaged $\overline{q}_A$  for  $d=0$ (normalized so that $\int_0^1 P(q_A)dq_A=1$).  Only the gap states are included in this analysis.   The parameters are $N_D=1000$, $n_{imp}=0.1$, $w=7.1$ \AA,  and $L=125.4$ \AA.  The histogram interval is $\Delta q_A=0.025$.}
\label{Prob0}
\end{center}
\end{figure}

Now, we examine the distributions of these  Type-I, -II, and -III states denoted by  $P(q_A)$, which is the probability distribution function showing the number of  {\it left and right}  gap-edge states having  a $q_A$ value in the interval $[q_A-\frac{\delta q_A}{2},q_A+\frac{\delta q_A}{2}]$.  As shown in Fig.~\ref{Prob0}, $P(q_A)$  is sharply peaked near $q_A=1/2$ for small $\Gamma$  and $d=0$.
Let us explain qualitatively  why the $P(q_A)$ of the gap-edge states is peaked at $q_A=1/2$ in the weak disorder regime.
This peak  implies  that 
there are numerous Type-II states in the gap in the presence of disorder. 
Two factors  are important  for this effect: (a) When $U\neq 0$ and $\Gamma=0$, the band structure calculation shows that the states near  the edges of the Brillouin zone $k=-\frac{\pi}{a_0}$ and $k=\frac{\pi}{a_0}$   are  {\it zigzag edge states}   (see Fig.~\ref{Ubandst1}); and (b)   
a short-ranged disorder potential  with $d\sim a_0$ couples these zigzag edge  states $\phi_{R,\uparrow}$ and $\phi_{L,\uparrow}$ near $k=-\frac{\pi}{a_0}$ and $k=\frac{\pi}{a_0}$, as
a significant wave vector transfer occurring in a backscattering process    is   $|k - k'|\sim 1/a_0$\cite{Lima}. 
Moreover,   when the disorder is weak, the DOS near the gap edges $E=\pm \Delta/2$ is {\it sharply peaked}.  
Because of this  sharp peak,   even a weak  disorder potential can mix the  Type-I left and right zigzag edge   states   and generate Type-II states with energy in the gap.

\begin{figure}[!hbpt]
\begin{center}
\includegraphics[width=0.3\textwidth]{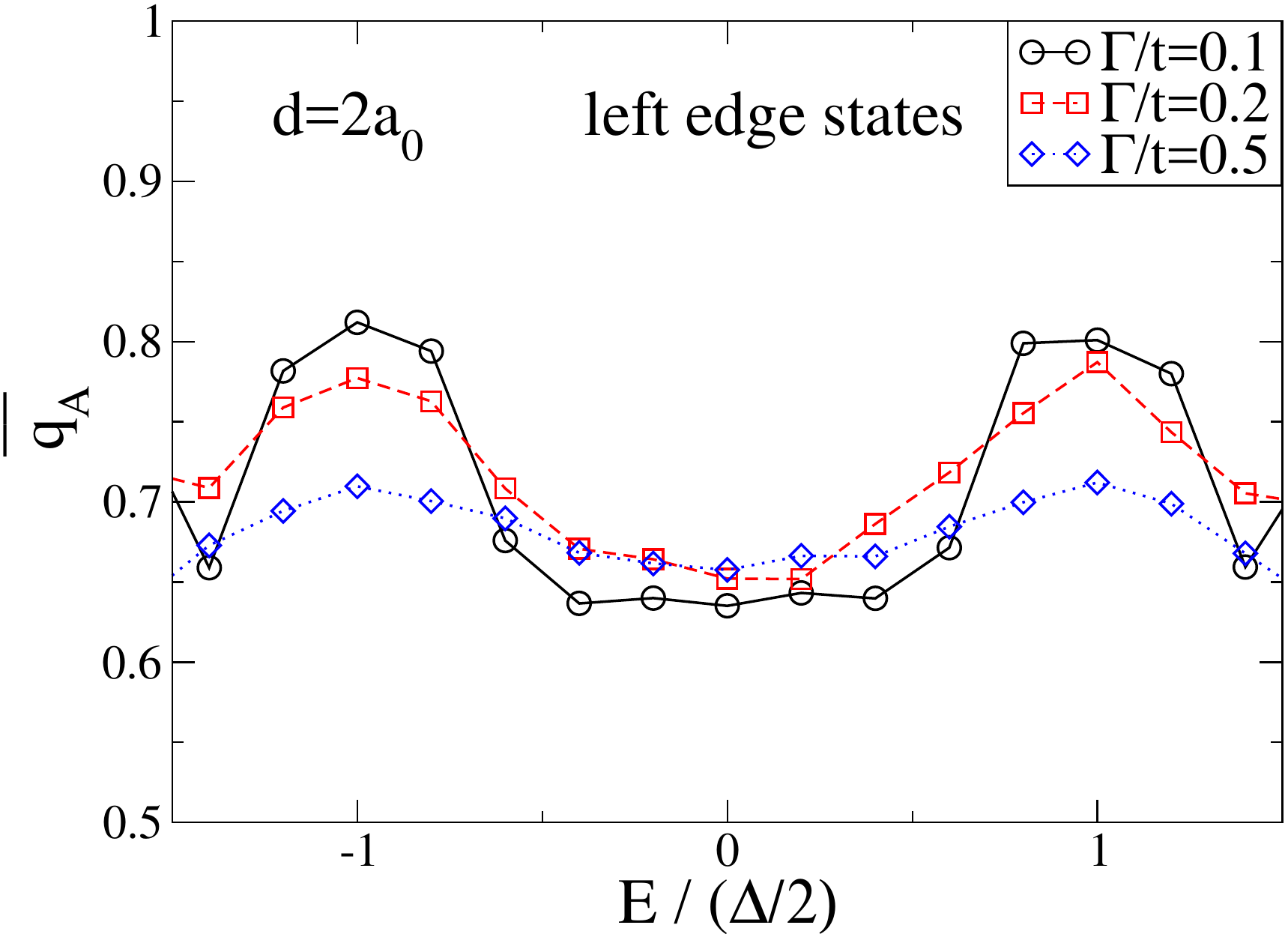}
\caption{ As in Fig.~\ref{qA0}, but for $d=2a_0$.  }
\label{qA2}
\end{center}
\end{figure}

\begin{figure}[!hbpt]
\begin{center}
\includegraphics[width=0.3\textwidth]{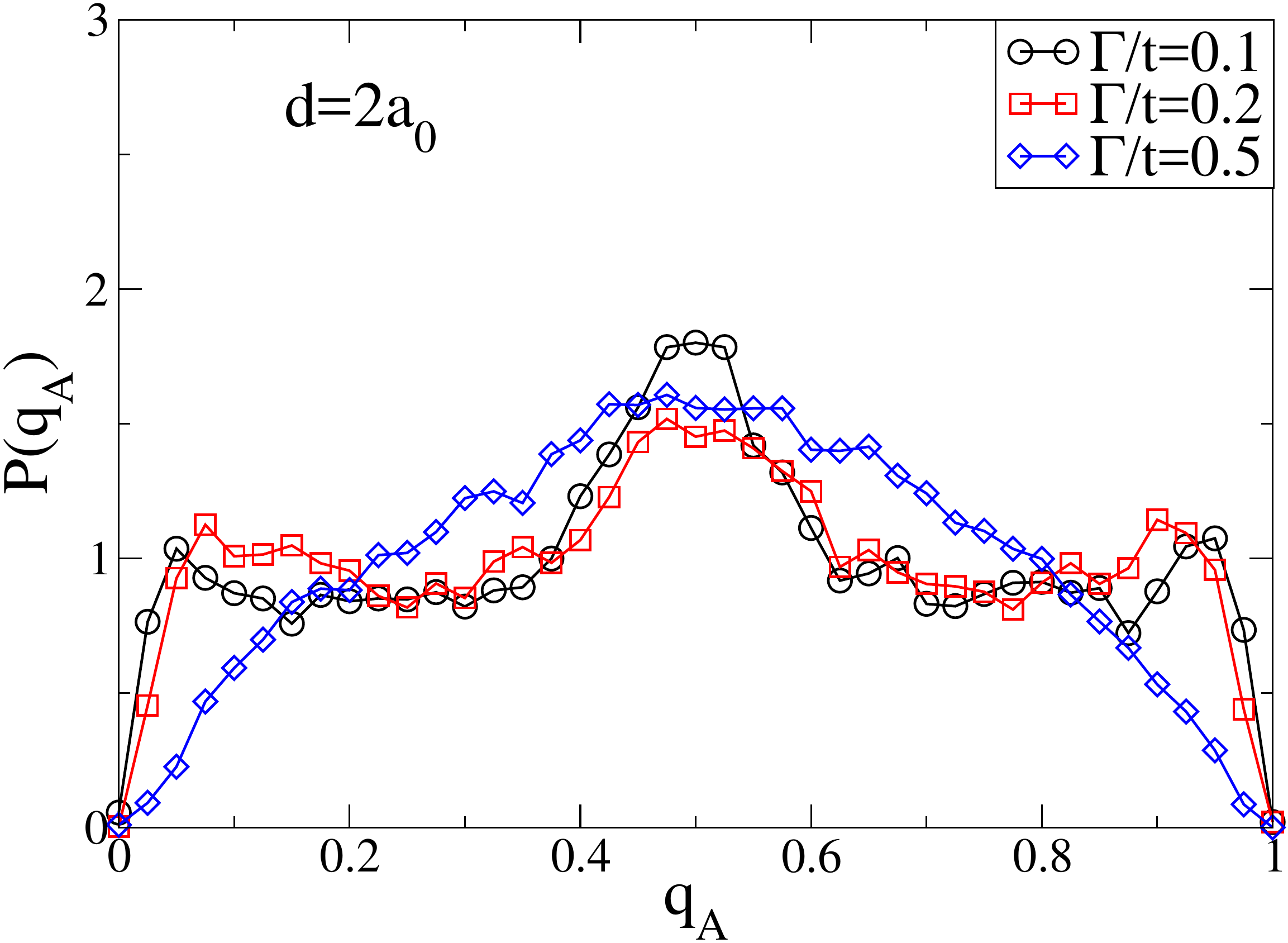}
\caption{ As in Fig.~\ref{Prob0}, but for $d=2a_0$.  }
\label{Prob2}
\end{center}
\end{figure}

The  $\overline{q}_A$ and $P(q_A)$ for $d=2a_0$ are shown in Figs.~\ref{qA2} and \ref{Prob2}, respectively
We see from Fig.~\ref{qA2} that, for small values of $\Gamma$ and $E\approx 0$,
the   $\overline{q}_A$ value of the left edge states is  not as  close to the quantized value $1/2$ as in the csae of $d=0$.   
Fig.~\ref{Prob2} shows $P(q_A)$ for the left and right edge states  combined.  Note that,   near $q_A=1/2$,  this function is not as sharply peaked as in the $d=0$ case.     

\section{Variance of fractional charge}

The previous result suggests that the midgap states may display quantization of the fractional charge.
The energy dependence of  $\overline{q}_A$
suggests that, if an electron with spin $\sigma$  is added into or removed from a midgap state in the energy interval $[-\frac{\delta E}{2},\frac{\delta E}{2}]$, it can become fractionalized into two $1/2$ charges located on left and right zigzag edges. However, to qualify for charge fractionalization, the variance of $q_A$ due to disorder must be negligible. Therefore, we here perform disorder averaging to compute the  variance 
of $q_A$.

\begin{figure}[!hbpt]
\begin{center}
\includegraphics[width=0.3\textwidth]{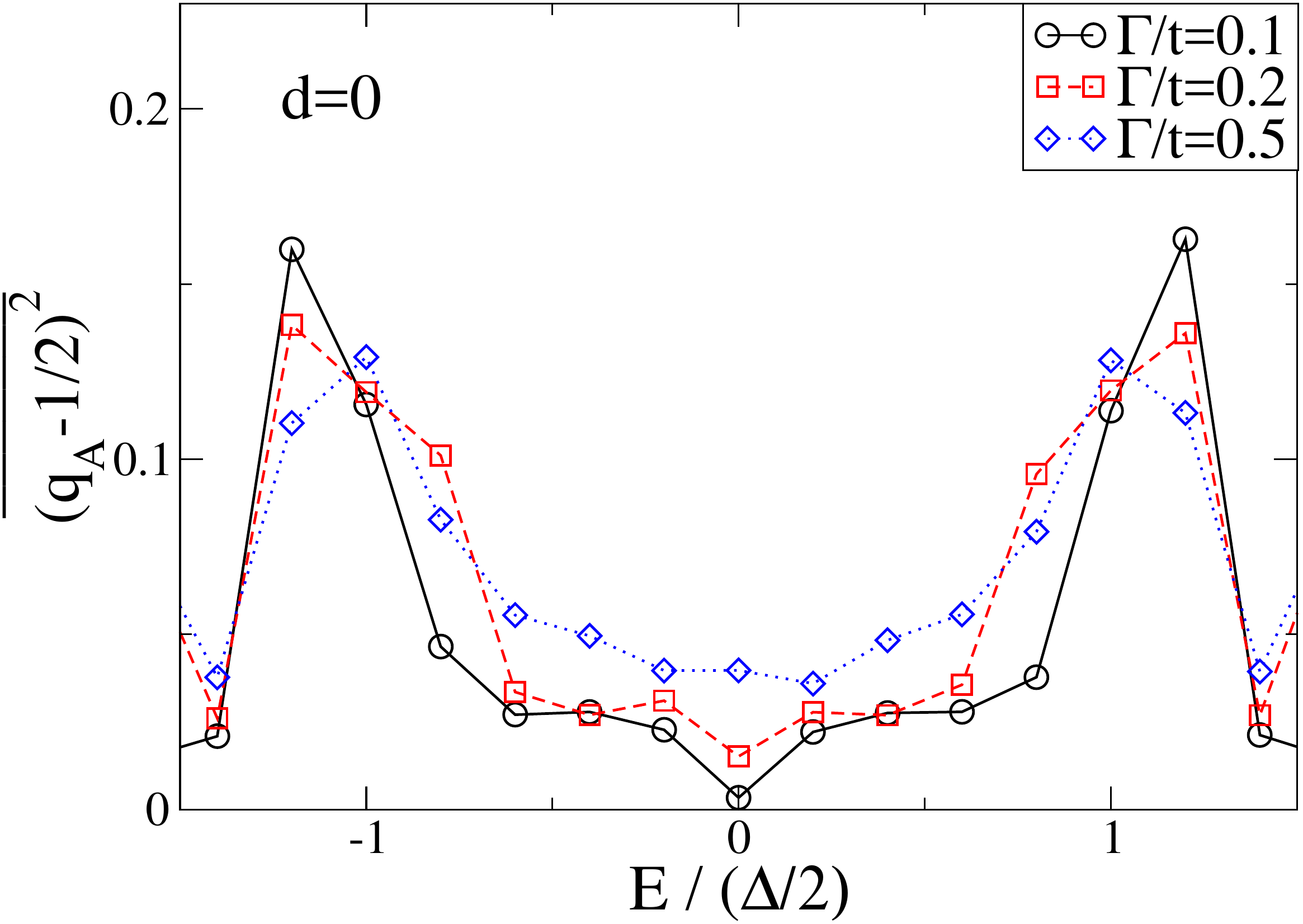}
\caption{Variance as  function of $E$  for  $d=0$. The  parameters are $N_D=1000$, $n_{imp}=0.1$, $w=7.1$ \AA, and $L=125.4$ \AA. Furthermore, 
$\delta E=0.1\Delta$.}
\label{varqA0}
\end{center}
\end{figure}

\begin{figure}[!hbpt]
\begin{center}
\includegraphics[width=0.3\textwidth]{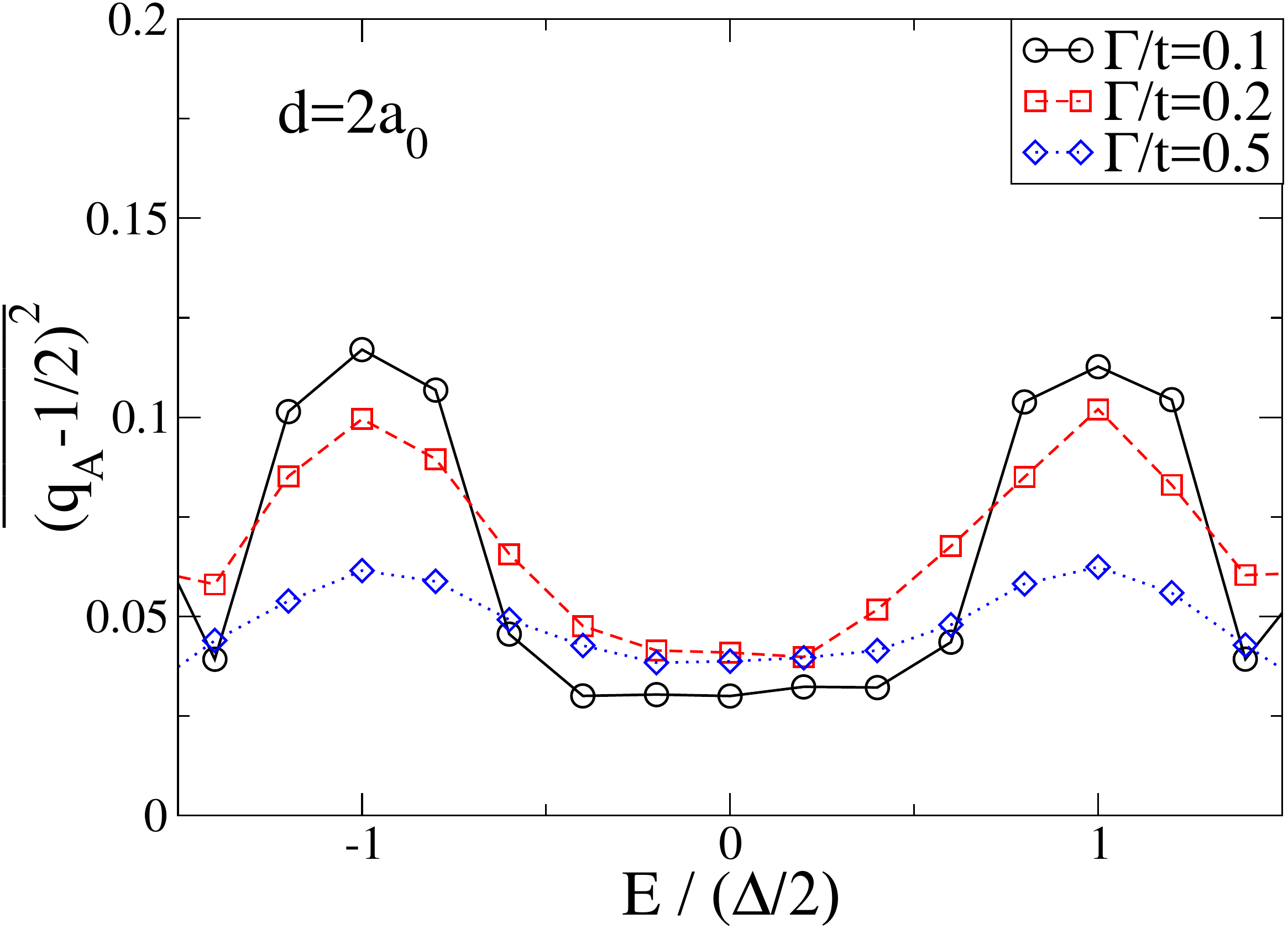}
\caption{Variance as  function of $E$  for $d=2a_0$. The other  parameters are as in Fig.~\ref{varqA0}.   }
\label{varqA2}
\end{center}
\end{figure}

\begin{figure}[!hbpt]
\begin{center}
\includegraphics[width=0.35\textwidth]{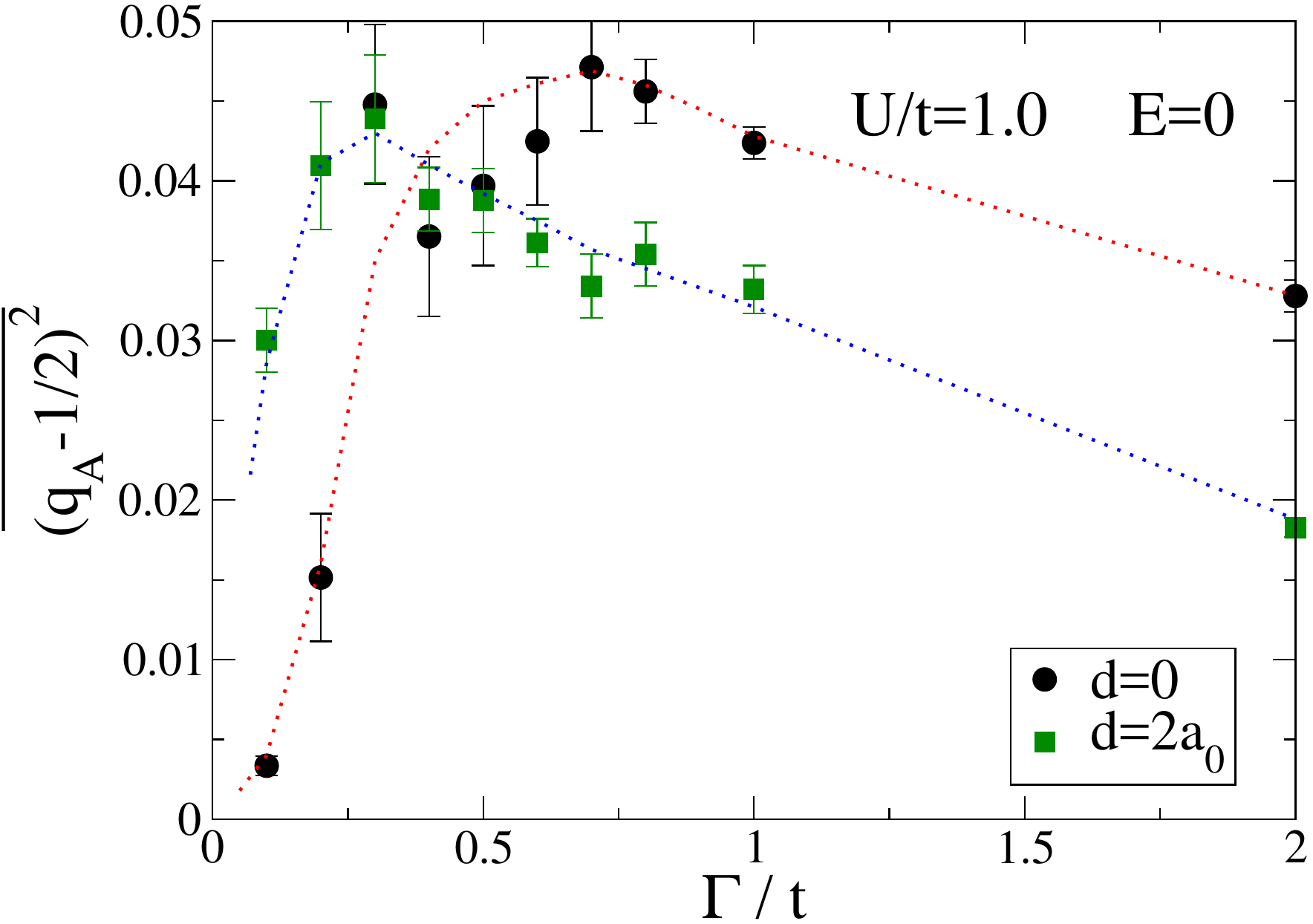}
\caption{Variance at  $E=0$ as a function of $\Gamma$ at $n_{imp}=0.1$ for  $d=0$ and $2a_0$ (the dotted line is a guide for the eye).  The other  parameters are the same as in Fig.~\ref{varqA0}.}
\label{varqA}
\end{center}
\end{figure}

\begin{figure}[!hbpt]
\begin{center}
\includegraphics[width=0.35\textwidth]{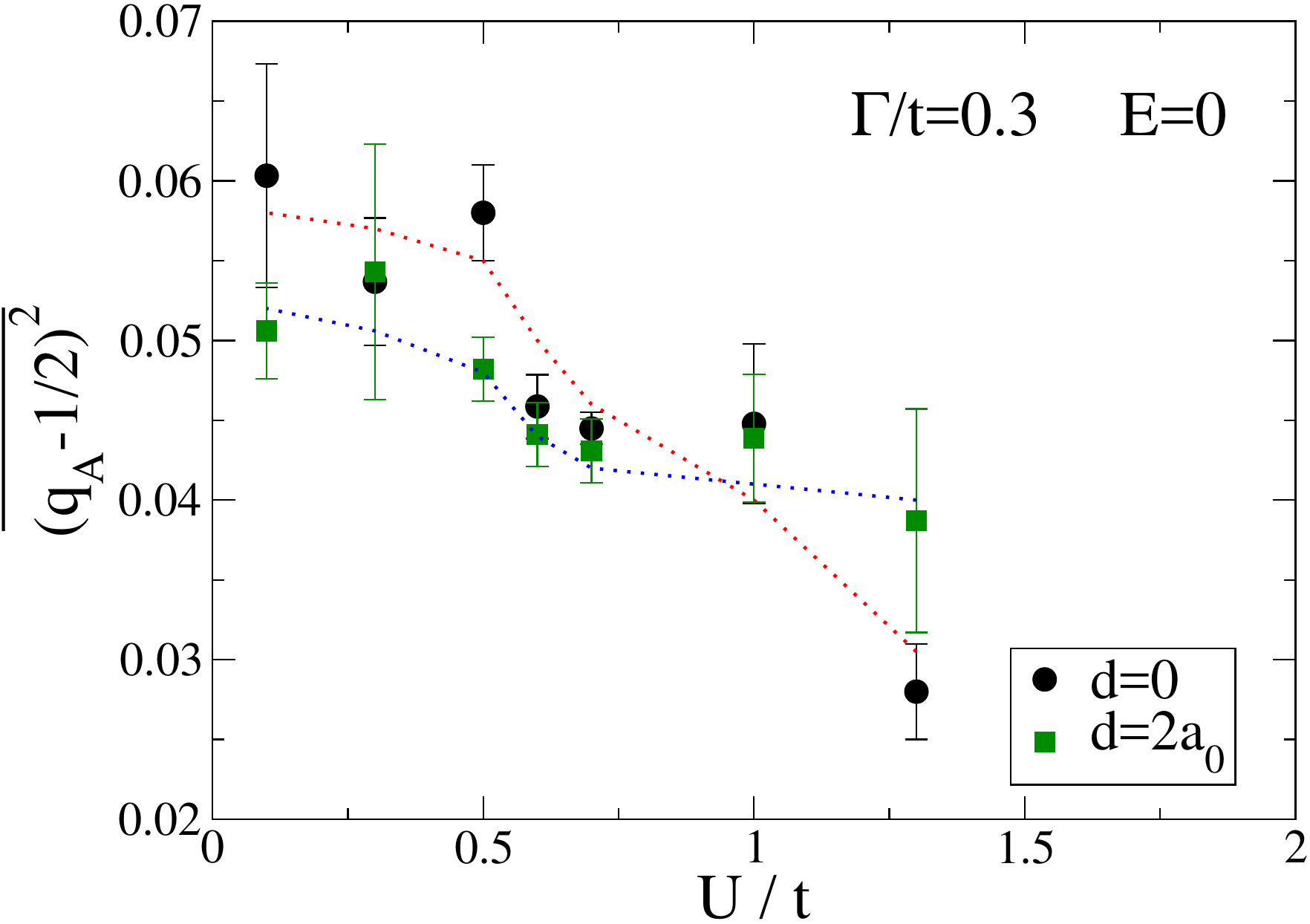}
\caption{Variance at  $E=0$ as  function of $U$ at $n_{imp}=0.1$ for  $d=0$ and $2a_0$ (the dotted line is a guide for the eye).  The value of $\Gamma$ is $0.3t$.  The other  parameters are as in Fig.~\ref{varqA0}.}
\label{varqAU}
\end{center}
\end{figure}

The variances $\overline{(q_A-\frac{1}{2})^2}$   for $d=0$ and $2a_0$ are shown in Figs.~\ref{varqA0} and \ref{varqA2}, respectively.  This behavior is independent of the energy interval $\delta E$ for sufficiently small  $\delta E$.   The variance is at a minimum for the midgap states. 

We also investigate the variance of  the midgap states for different values of $\Gamma$ and $U$.
Fig.~\ref{varqA} displays the variance   as a function of $\Gamma$ at a fixed value of $U=t$ while 
Fig.~\ref{varqAU} displays the variance  as a function of $U$ at a fixed value of $\Gamma=0.3t$. These figures show that the charge variance of the midgap states is small
in the weak disorder regime $\kappa={0.33\Gamma}/{U}\ll 1$  $(with \sqrt{  n_{imp}  }=0.33)$.   
The variance reduces rapidly as $\Gamma $ decreases:     for $d=0$,  its
magnitude   is  reduced to $0.004$ at $\Gamma=0.1t$. 
In the weak disorder regime, the variance of the $1/2$ fractional charge is smaller for  $d=0$ than for  $d=2a_0$.  However, when $\kappa\gtrsim 1$, the opposite is true.

The results of Figs.~\ref{varqA} and \ref{varqAU} suggest that the variance tends to   decrease further when $\Gamma $ is reduced below $\Gamma=0.1t$ (note that {\it reducing}   the  value of $n_{imp}$
has similar effects to reducing $\Gamma$).  However, there are  fewer midgap states in the histogram interval (note that the density of states (DOS) near $E \approx 0$ is rather small).  To circumvent this problem, a larger value of the  ribbon length $L$ is needed to allow inclusion of   more
midgap states  in the histogram  energy interval (we have verified numerically that there are indeed more midgap states when the length is increased to $L=307.4$ \AA).  However, disorder averaging and computation of the variance for such a system 
is numerically demanding.    Nonetheless, our numerical work indicates that, when  the disorder potential is weaker than the on-site repulsion,   the charge fractionalization of the midgap state is robust.   We believe that the accuracy of the fractional charge quantization in the limit $\kappa\rightarrow 0$ is related to the presence of the particle-hole symmetry of the underlying band  structure in the absence of disorder (the particle-hole symmetry also plays an important role  in polyacetylene\cite{Sol}).  We do not expect  that the fractional charge quantization is precise 
for long-ranged disorder potentials with $d\gg a_0$, as the disorder-induced coupling between the left and right edge  states is reduced in comparison to that of short-ranged potentials.

\section{Excitation gap and quantized charge}

\begin{figure}[!hbpt]
\begin{center}
\includegraphics[width=0.4\textwidth]{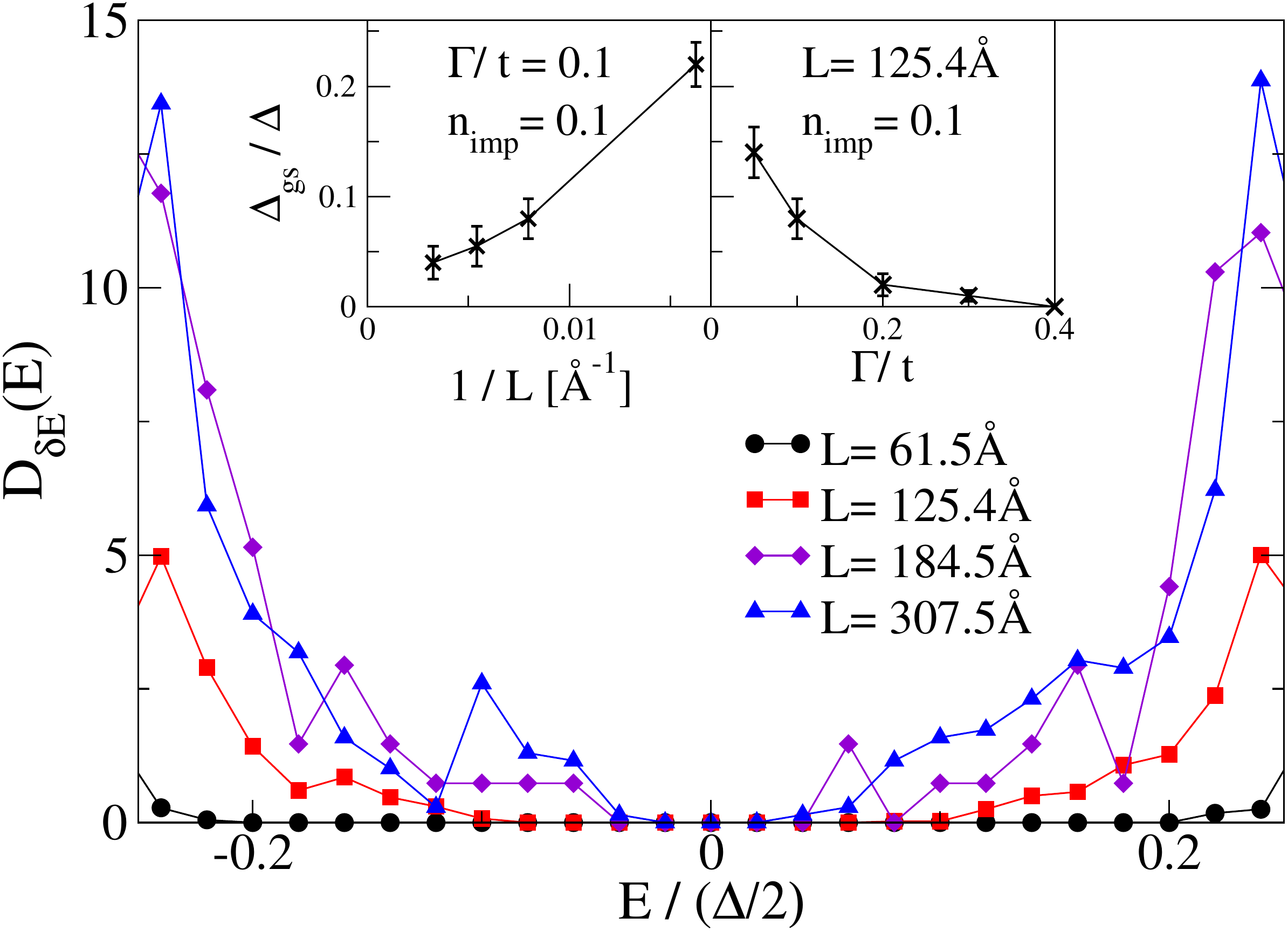}
\caption{Number of states $D_{\delta E}(E)$ in the energy interval $\delta E$ is shown  for several values of $L$.  Note that there is an excitation gap near $E=0$.   Finite-size scaling result for the excitation gap $\Delta_{gs}=\Delta(L,\Gamma\sqrt{n_{imp}},U)$ is shown in the left  inset.  The right inset displays how $\Delta_{gs}(L,\Gamma\sqrt{n_{imp}},U)$ vanishes as $\Gamma $ increases. Here, $\delta E=0.01\Delta$.   The other  parameters are as in Fig.~\ref{varqA0}.}
\label{DOS}
\end{center}
\end{figure}

To qualify for a true charge fractionalization the quantum charge fluctuations should takes place at high frequencies.  According to Girvin\cite{Girvin},  the characteristic time scale for the charge fluctuations is inversely proportional to the relevant excitation gap.  As we mentioned in the previous section, there are only few midgap states in the energy interval $\delta E$ near $E=0$.  This is because there is a small gap $\Delta_{gs}(L,\Gamma\sqrt{n_{imp}},U)$ between the occupied and unoccupied midgap states in the weak disorder regime, see Fig.\ref{DOS}.  The magnitude of this excitation gap depends on the  length of the zigzag edges $L$, the
disorder strength $\Gamma\sqrt{n_{imp}}$ and interaction strength $U$.  $\Delta_{gs}(L,\Gamma\sqrt{n_{imp}},U)$ increases approximately linearly with $U$, see Ref.\cite{Jeong0}.  The excitation gap  depends on  $L$   because it is the zigzag edges that generate 
topological edge states and the total number of edge states depend on $L$\cite{Jeong0}.  
We find that  this excitation gap is larger for weaker disorder strength (note also that the bandgap   $\Delta_{gs}(L,0,U)$ develops in the DOS in the limit of zero disorder strength).   On the other hand, no gap exists when disorder is strong.
We have estimated the size of the gap numerically at $L=307.4$ {{\AA}}:  we find $ 0.05\Delta\sim 10^{-2}t\sim 10 \textrm{THz}$; here disorder strength $\Gamma=0.1t$  and edge separation  $w=7.1$\AA).  This value corresponds to very a short time scale.

Now we discuss the limit $L\rightarrow \infty$.   It is difficult to compute quantitatively the size of the excitation gap  in this case since our numerical approach is limited to systems with $L\lesssim 300$\AA.
Our available numerical result  suggests  that, in this limit, 
$\Delta_{gs}(L,\Gamma\sqrt{n_{imp}},U)$ may not vanish in the weak disorder regime, see Fig.\ref{DOS}.     However, results for significantly longer values of $L$ are needed to establish this. If a pseudogap\cite{Eric,Gun}  exists in the limit $L\rightarrow \infty$ charge quantization may  be only approximate.

\section{Summary}

In this study, we investigated the  properties   of the gap-edge states of half-filled interacting disordered  zigzag graphene nanoribbons (ZGNRs).    In disorder-free GNRs, antiferromagnetic coupling between well-separated  zigzag edges is detrimental to the formation of fractional edge charges.  However, disorder may increase  the coupling between the edges, which may favor the formation of midgap states with a fractional edge charge.  Our work shows that  the disorder plays an important role in mitigating  the effect of antiferromagnetic coupling on  the midgap states.    The coupling of the left and right edge states by a short-ranged potential is important for this effect.  As shown in Fig.~\ref{pseudo}, the site pseudospin of a midgap state connects the left and right zigzag edges with different directions of magnetization.  A soliton fractional charge is thus protected topologically against  weak disorder.

Using a self-consistent Hartree-Fock mean field approach, we 
found that the variance of the fractional charge of the midgap states is determined by the competition among the disorder strength $\Gamma$,   impurity range $d$, and on-site repulsion $U$.   In the weak disorder regime, the ratio between the disorder and the interaction strengths, $\kappa\ll 1$,  and for short-ranged potentials,  the charge variance of these midgap states is small and  a well-defined  quantized fractional charge of $1/2$   can be formed.  Our numerical result suggests that, in the weak disorder regime, the quantized fractional charge becomes more  precise as the values of   $\Gamma$, $n_{imp}$ (the ratio between the number of impurities and the total number of carbon atoms), and $d$ decrease.  We believe that  the accuracy of the fractional charge quantization in the weak disorder limit $\kappa\rightarrow 0$ is related to the particle-hole symmetry of the underlying bandstructure in the absence of disorder.  For longer-ranged impurities,  the variance is larger and  decreases more slowly  as $\kappa\rightarrow 0$. 
Other gap states with $E\neq 0$ have larger charge fluctuations than those of the midgap states.   In addition,  we have found that numerous gap-edge states become  spin-split and singly occupied (note that the Mott-Anderson insulator also has spin-split singly occupied states).

Our numerical study indicates that charge quantization may be realized under the following conditions in finite length ribbons.  First, an excitation gap must be present.  This insures that the  midgap states display small  quantum charge fluctuations.   Second,  midgap states must be close  to $E=0$. This condition is satisfied when the excitation gap is  small in comparison to the bandgap, $\Delta_{gs}(L,\Gamma\sqrt{n_{imp}},U)<<\Delta_{gs}(L,0,U)$.   At the same time we must require  that the variance due to disorder is small, which is satisfied  in
the weak disorder regime  $\kappa<< 1$.   When   $\kappa>>1$ disorder  destroys the excitation gap   and charge quantization is not expected.   In the limit $L\rightarrow \infty$,  our numerical results suggest that a finite excitation gap may exist, but results for significantly longer values of $L$ are needed to establish this with certainty.      Instead of performing numerical work for larger values of $L$, it may be more promising to develop a field theoretical model in the presence of disorder\cite{Jeong0,Jack2}.  If a pseudogap exists in the limit $L\rightarrow \infty$ we expect the charge quantization to be less accurate.

Considerable experimental effort has been expended to measure the edge states in ZGNRs. Measurement of the differential conductance in atomically precise ZGNRs\cite{Ruff,Hao} using  scanning tunneling microscopy\cite{Andrei} may provide rich information on the edge charge variations.   The measurement  of the distribution of type I, II, and III gap states as a function of energy and disorder strength is desirable.

\section*{Acknowledgments}
This research was supported by the Basic Science Research Program
through the National Research Foundation of Korea (NRF), funded by the
Ministry of Education, ICT $\&$ Future Planning (MSIP) (NRF-2018R1D1A1A09082332  to S.R.E.Y. and NRF-2016R1D1A1B03935815 to M.C.C.).

\appendix

\section{Magnetic properties of  gap-edge states}

Here, we investigate the manner in which the gap-edge states  contribute to the
magnetic  properties of ZGNRs.  The edge magnetism decreases as the disorder parameter $\Gamma$ changes.  
Quantitative analysis of the magnetization at the $i$th site  may be performed by averaging over many disorder realizations.     The disorder-averaged value of the site magnetization is defined by
\begin{equation}
\overline{| \langle s_{i}\rangle_{\Gamma}|}= \overline{   |\langle n_{i\uparrow}\rangle  - \langle n_{i\downarrow}\rangle|} ,
\label{spin}
\end{equation}
where the overline indicates  a disorder-averaged value.  Note that  the Type-I, -II, and -III
gap-edge states contribute to the occupation numbers $\langle n_{i\sigma}\rangle$,  in addition to the non-edge states (see Eq.~(\ref{occnum})).  However, the contribution from  the gap-edge states is dominant.

\begin{figure}[!hbpt]
\begin{center}
\includegraphics[width=0.35\textwidth]{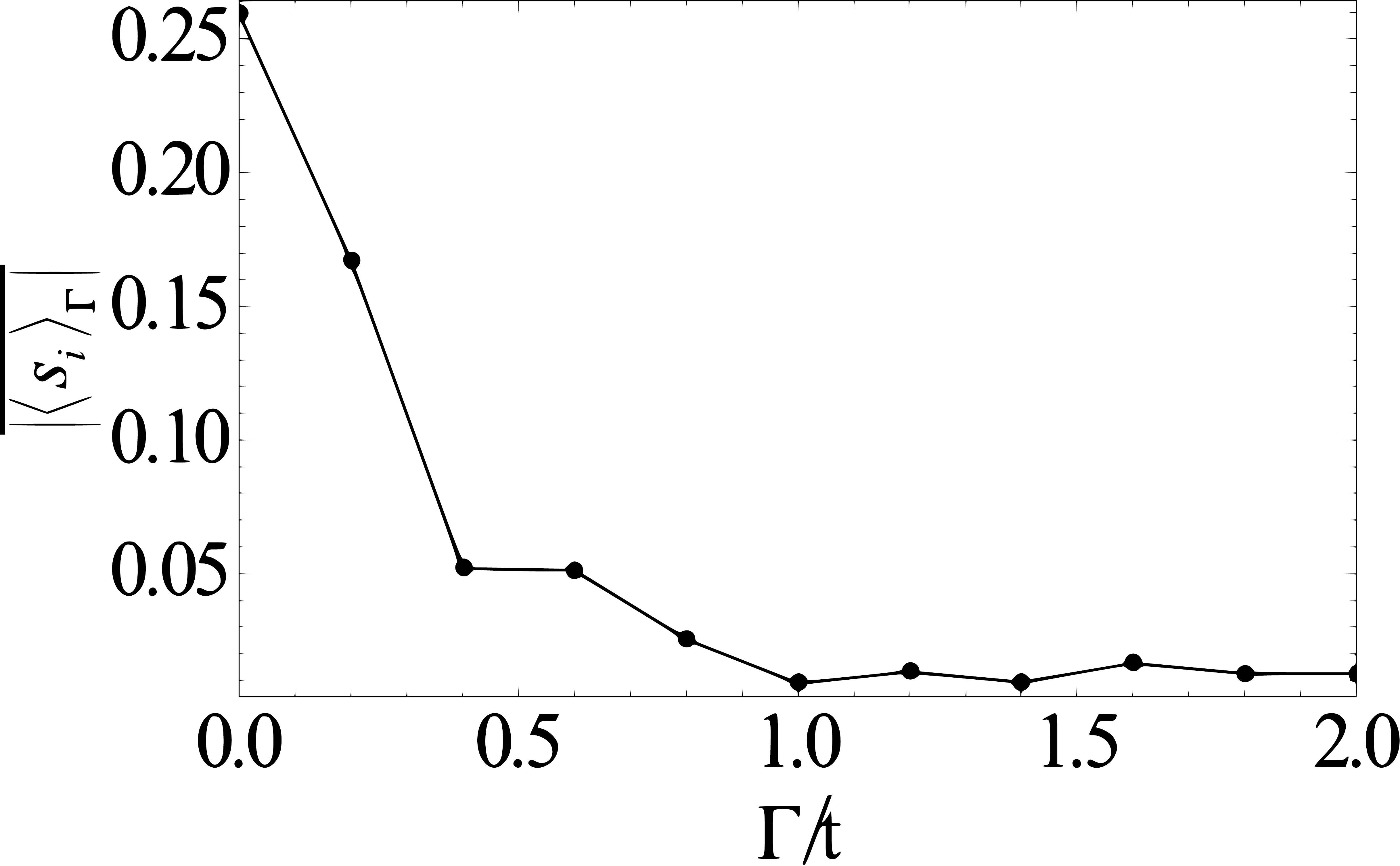}
\caption{ Disorder-averaged spin value per edge site vs. $\Gamma$. The parameters are   $d=0$, $n_{imp}=0.1$, $L=125.4$ \AA, and $w=7.1$ \AA. Furthermore, $N_D=10$.    }
\label{edgemag}
\end{center}
\end{figure}

In Fig.~\ref{edgemag}, $\overline{|\left\langle s_{i}\right\rangle_{\Gamma}|}$ is plotted as a function of  $\Gamma$.   At $\Gamma=0$, the Type-I edge states contribute to a finite edge magnetization and
the expectation value of the edge magnetization per site on the boundaries is $\langle s_{i}\rangle=\langle n_{i\uparrow}\rangle-\langle n_{i\downarrow}\rangle=\pm 0.24$. The value $\langle s_i\rangle $ decays into the ZGNR over several carbon-carbon distances.
However, along the zigzag edges, $ \langle s_{i}\rangle$ is uniform.  The edge magnetization decreases with increasing $\Gamma$.
Our results show that, in addition to the proliferation of Type-II gap-edge states (topological kinks), the partial spin polarization of the Type-I and -III gap-edge states on each zigzag edge plays a role in this reduction.   
Note that, although it is small, the edge antiferromagnetism persists even in the presence of a substantial amount of disorder.


\begin{thebibliography}{9}
\bibitem{pseudo}   Consider a graphene sublattice basis, consisting of two carbon atoms A and B.  Suppose that the probability amplitudes of finding an electron on the A and B atoms are $c_1$ and $c_2$, respectively.   
This can be represented as a spinor $\left(\begin{array}{c}
c_1\\
c_2
\end{array}
\right)$.  Using this spinor, the components of the pseudospin  are defined as  the expectation values of the Pauli spin matrices $\sigma_x,\sigma_y,\sigma_z$.
\bibitem{def} A kink is a one-dimensional topological object with a phase change of $\pi$ when the position coordinate varies through the kink (the crucial feature is that the manner in which the phase changes as the coordinate changes is irrelevant).   In our case, the phase is the angle of the pseudospin vector.
\bibitem{com} In graphene systems, only a  short-ranged disorder potential is effective in inducing localization.
Depending on the electron energy, perfectly conducting channels   may appear\cite{Lima}. These properties are in sharp contrast to those of ordinary one-dimensional localization theory.
\end{thebibliography}
\end{document}